\documentclass{article}

\usepackage{arxiv}

\usepackage[T1]{fontenc}
\usepackage[utf8]{inputenc}
\usepackage[english]{babel}
\usepackage{graphicx}
\usepackage{amsmath}
\usepackage{amsfonts}
\usepackage{hyperref}
\usepackage{fancyhdr}
\usepackage{lmodern}
\usepackage{cancel}
\usepackage{amssymb}
\usepackage{amsthm}
\usepackage{pgfplots}
\usepackage{accents}
\usepackage{bbm}
\usepackage{bm}
\usepackage{esint}
\usepackage{mathrsfs}
\usepackage{booktabs}
\usepackage{xcolor}
\usepackage{gensymb}

\newcommand{\wt}[1]{\widetilde{#1}}
\newcommand{\wh}[1]{\widehat{#1}}

\theoremstyle{plain}

\theoremstyle{definition}

\title{Interfacial instability as a trigger for dryout inception in two-phase CO\textsubscript{2} flow} 

\author{  
G. Cantini \href{}{\includegraphics[scale=0.1]{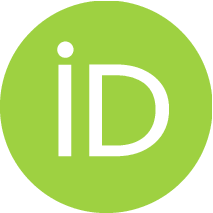}} \\ \small Department of Mechanical Engineering \\ University of Bath \\ Claverton Down, Bath BA2 7AY \\ United Kingdom \\ 
\texttt{gc803@bath.ac.uk} \\
\And 
G. Arnone \href{}{\includegraphics[scale=0.1]{orcid.eps}} \\ \small Istituto Nazionale di Alta Matematica “F. Severi” \\ INdAM, Piazzale Aldo Moro 5, Roma, 00185 \\ Italy \\ 
\texttt{arnone@altamatematica.it} \\
\And
F. Capone \href{}{\includegraphics[scale=0.1]{orcid.eps}} \\ \small Dipartimento di Matematica e Applicazioni R. Caccioppoli \\ University of Naples Federico II \\ Napoli, 80126\\ Italy \\ 
\texttt{fcapone@unina.it} \\
\And
J.A. Gianfrani \href{}{\includegraphics[scale=0.1]{orcid.eps}} \\ \small Dipartimento di Matematica e Applicazioni R. Caccioppoli \\ University of Naples Federico II \\ Napoli, 80126\\ Italy \\ 
\texttt{jacopoalfonso.gianfrani@unina.it} \\
\And
M. Carnevale \href{}{\includegraphics[scale=0.1]{orcid.eps}} \\ \small Department of Engineering and Applied Sciences \\ University of Bergamo \\ Via Marconi 4, 24044 Dalmine (BG)  \\ Italy \\ 
\texttt{mauro.carnevale@unibg.it}
}

\renewcommand{\shorttitle}{}

\begin{document}
\maketitle
\begin{abstract}
Progress in particle physic leads to increasing in detector luminosity and a consequent increasing overheating induced by Joule effect. An effective cooling strategy is the exploitation of CO\textsubscript{2} heat latency in phase-change. An additional challenge, relevant to detectors for High Energy Particles, is the consequent geometrical constrain due to the limited space avialable for the cooling system within the detector arrangement, leading to the implementation of cooling system by means of millichannels. In this context, at relative high vapour quality the liquid phase exhibits annular flow, anticipating the dryout. Dryout is a critical condition where the heat transfer coefficient dramatically drops and dangerous temperature levels can be reached, potentially leading to catastrophic consequences.
Experimental evidences reveal that its behavior in two-phase annular flows differs from conventional refrigerants and the fundamental inception-mechanism is not yet understood.
This study aims at investigating the key new idea whereby dryout inception is triggered by instability of the liquid-vapour interface. A mathematical model for two-phase annular flow is presented and the stability of the interface between the two fluids is studied through the linear theory. The stability analysis reduces to solving a coupled forth-order differential eigenvalue problem that is treated numerically with an in-house code based on the Chebyshev-$\tau$ method. Numerical investigations identify a critical value for the vapour quality, named $x_{dry}$, that leads to interface instability. The resulting predictions on $x_{dry}$ are confirmed by experimental data collected from two independent experimental campaigns, validating the hypothesis that dryout inception is governed by interfacial instabilities.
\end{abstract}

\section{Introduction}

The dryout phenomenon is known in heat transfer processes as a condition to be avoided in order to keep a cooling system working at its best \cite{weislogel1998hydrodynamic}. It is defined as the condition where the Heat Transfer Coefficient (HTC) drops drastically within the heat exchanger, reaching values comparable to those of gaseous flows \cite{bergman2011introduction}. In heat management processes, during the evaporation the flow is bi-phase and the heat transfer is more efficient than in flows without phase change \cite{collier1994}. Nonetheless, as the vapour quality increases along the channel, the progressive depletion of the wall film may trigger dryout. In such a condition, the liquid film detaches from the pipe wall that is then in direct contact with the vapour phase, leading to an abrupt degradation of the local HTC and potentially compromising component integrity.
In applications where there is a constant imposed heat flux at the wall, like in nuclear reactor or in electronics cooling, the dryout can raise the wall temperature enough to melt the exchanger bringing the devices to failure or dangerous consequences \cite{hewitt2013annular}. Therefore, a more accurate prediction of the onset of dryout can be useful to support less conservative safety margin in cooling system design, with a more efficient result.

Among all refrigerants, boiling carbon dioxide offers numerous advantages over synthetic refrigerants. Its cost, reliability and minimal Global Warming Potential make CO\textsubscript{2} cheap, efficient and environmentally friendly. The high working pressure and low liquid-to-vapour density ratio of CO\textsubscript{2} make it well-suited for efficient heat exchange in milli-scale diameter pipes and for saturation temperatures $T_{sat}$ in the interval $(-40^\circ C,-10^\circ C)$ \cite{gullo2017state}. This is because the evaporation takes place at higher pressure than standard refrigerants and the pressure drop has a limited relative effect on boiling pressure compared to standard fluids \cite{Cern}.
Additionally, CO\textsubscript{2} is neither toxic nor flammable, and its non-corrosive properties are pivotal for safe operation of the electronics and other equipment. Its resistance to radiation makes it an ideal candidate for cooling applications in particle physics detectors. Furthermore, for small-diameter channels the thermal conductivity of CO\textsubscript{2} is higher than
standard fluids as can be seen in Figure 2 in \cite{Cern}.

The above features, however, make CO\textsubscript{2} two-phase flows qualitatively different from those of conventional refrigerants. As a result, phenomenological models struggle to accurately predict the vapour quality at the onset of dryout, from now on denoted as $x_{dry}$. Accurate estimation of $x_{dry}$ is therefore pivotal in cooling-system design. In \cite{revellin2008conditions}, two trends have been identified for the CO\textsubscript{2} dryout vapour quality as a function of the mass flux:
\begin{itemize}
    \item  {$\mathbf{\delta}^+$ regime}: $x_{dry}$ increases with rising mass flux $G$. Although less frequently observed in the literature, some instances of this behaviour are documented in \cite{YUN20032527} and \cite{ducoulombier2011carbon}.
    \item {$\mathbf{\delta}^-$ regime}: $x_{dry}$ decreases with increasing mass flux $G$. This regime is widely reported in the literature as in \cite{wojtan2005a} and \cite{cheng2008a}.
\end{itemize}
The flow pattern map proposed by \cite{cheng2008a} is considered the most robust for CO\textsubscript{2} but represents solely $\delta^-$ behaviour. Correlations informed by the $\delta^-$ regime will result in an underestimation of $x_{dry}$ when $\delta^+$ behaviour is present. An attempt to describe the transition from $\delta^+$ to $\delta^-$ is proposed in \cite{revellin2008conditions} where a phenomenological model based on a system of differential equations for liquid and vapour phases is introduced.   However, such a model  disregards the wall heat flux, 
which has been shown to be important in the $\delta^+$ regime through an experimental study conducted in \cite{ducoulombier2011carbon}. Importantly, all theoretical approaches presented in the open literature fail to model the compound effects of all contributing physical parameters, as stressed by \cite{hellenschmidt2021effects}.

A recent paper \cite{cantini2025inception} provides an initial step toward filling this gap. This research is based on an experimental campaign conducted at CERN, and is intended to investigate the dryout inception in CO\textsubscript{2} in milliscale pipes under representative operating conditions for high-heat-flux applications in the $\delta^+$ regime.
In \cite{cantini2025inception} the authors show the inability of existing models to predict dryout inception across all conditions, particularly the less understood $\delta^+$ regime  where dryout vapour quality increases with mass flux. Moreover, 
as a novelty, the authors connect experimental observations to stability theory through dimensional analysis, deriving a \textit{dryout instability factor}, denoted as $I_{\delta^+}$, that naturally emerges from momentum balance at the interface. The empirical correlation between $I_{\delta^+}$ and measured $x_{dry}$ bridges first-principles theory and data-driven prediction, validated against independent data from \cite{ducoulombier2011carbon}. 
Remarkably, this framework is independent of the saturation temperature and the heat flux, namely it reduces the dependence on experiment-specific conditions, making design and extrapolation easier.

In this context, motivated by \cite{cantini2025inception}, the present work aims to support the hypothesis that the dryout phenomenon is triggered by instability of the interface between liquid and gaseous CO\textsubscript{2}, providing a mathematical theoretical counterpart of the experimental empirical campaign. To this end, we develop a mathematical model for two-phase annular flow, with liquid and vapour separated by a sharp interface with appropriately defined interfacial conditions. 

The present study represents the first application of rigorous stability theory to the $\delta^+$ regime in boiling CO\textsubscript{2} annular flow, providing a framework that simultaneously accounts for all relevant mechanisms: viscous effects, surface tension, heat and mass transfer,  and compound effects of geometry and operating conditions.
The resulting equations form a differential eigenvalue problem of a couple of fourth-order differential equations requiring eight boundary conditions: four at the interface, two at the pipe centerline, and two at the wall.

We show that, for a prescribed set of parameters, solving the eigenvalue problem bridges stability characteristics with $x_{dry}$, by identifying dryout inception when the real part of the leading eigenvalue becomes positive. 
Numerical results show good agreement with experimental data from two independent campaigns \cite{ducoulombier2011carbon,cantini2025inception}.
At this point, a natural question arises: is CO\textsubscript{2} the only fluid suitable to the present study, or does it realize, under practical operating conditions, the thermodynamic state required for the onset of the $\delta^+$ regime? The latter is the answer. Specifically, compared to most refrigerants, CO\textsubscript{2} exhibits much smaller liquid-to-vapour density ratio (i.e. more similar liquid and vapour densities) at typical refrigeration operating temperatures. A relatively small density ratio implies a reduced slip ratio between vapour core and liquid film, hence weaker interfacial shear and smoother interface in annular flow. These are precisely the conditions under which the interfacial-stability mechanism considered here becomes physically meaningful and predictive for dryout inception. Conversely, for standard refrigerants at the same temperatures, the much larger density ratio promotes $\delta^-$ behaviour. Actually, only outside the standard operating conditions relevant for evaporators (e.g. saturation temperatures
far above practical refrigeration operation), the $\delta^+$ regime dominates for standard refrigerants.
The here proposed model implicitly contains a  hypothesis that the $\delta^+$ regime might exist for other fluids at some conditions, but those conditions are not practically relevant. Only CO\textsubscript{2} can achieve  specific  thermodynamic state at practical operating conditions required for $\delta^+$ behaviour.
The reader is referred to Section \ref{tabelle} for details.


In literature, interfacial instability between two superposed fluids has been widely analyzed \cite{chandrasekhar1981hydrodynamic,drazin2004hydrodynamic,adham1995rayleigh,hsieh1972effects,hsieh1978interfacial}. In particular, \cite{hsieh1972effects}  proposed a model for interfacial instability between two superposed homogeneous compressible fluids at rest that are invested by thermal gradients. These fluids occupy a horizontal layer and the interface between them (identified by a \textit{horizontal} profile) divides the region in two parts and it is treated as a sharp interface. Therefore, interfacial conditions are appended to the model and the effect of evaporation and interfacial waves on the onset of Rayleigh-Taylor instability is studied. Later on, the same author proposed in \cite{hsieh1978interfacial} a simplified mathematical model of the same physical setting where thermal effects modify the occurrence of interfacial instability. The main assumption that simplifies the model regards the replacement of the energy balance equation with an equation of production of mass at the interface. This is relying on the fact that \emph{since the transfer of mass across the interface represents a transformation of the fluid from one phase to another, there is invariably a latent heat associated with the phase change}. The author then remarks that \emph{it is essentially through this interfacial coupling between the mass transfer and the release of latent heat that the motion of fluids is influenced by the thermal effects} \cite{hsieh1978interfacial}. In other words, the net heat flux across the interface is modeled as a function of interface displacement.
In the present study, we follow this idea.

Unlike the above-mentioned works, the physical framework examined here involves heat flux at the pipe wall that provides external heat to the system and cylindrical geometry constraint. A fundamental assumption of this research is to model the presence of the external heat flux via a non-constant profile of the interface over the axial length of the pipe. For this reason, the liquid film thickness is assumed to be reducing along the axis of the pipe, so as to model the heat and mass transfer from the liquid phase to the vapour one. Moreover, 
despite the model being defined upon two-dimensional approximation, it keeps track of the cylindrical geometry within the interfacial conditions, via the dryout instability factor $I_{\delta^+}$, defined upon relations that hold only in cylindrical geometries, see Eq. \eqref{cantinis}. In this regard, let us remark that the $\delta^+$ regime in CO\textsubscript{2} two-phase flow can be achieved only in millichannels where the annular flow occurs as larger geometries would promote gravitational stratification of the flow, potentially  shifting behavior toward $\delta^-$ even when density ratios remain favorable. Studies  using larger channels may not access the $\delta^+$ regime despite using CO\textsubscript{2}.



The plan of the paper is the following. In Section \ref{sec2}, the physical problem is introduced via a schematic framework with the aim of clarifying the geometry of the problem and introducing the relevant variables appearing in the mathematical model. In Section \ref{sec3}, the derivation of the mathematical model and interfacial conditions is provided, including the main assumptions that lead to a mathematically treatable problem. The dimensional analysis is shown in Section \ref{sec4}, where the nondimensional dryout instability factor is defined.
Section \ref{tabelle} deals with the rationale for considering CO\textsubscript{2} in the present study instead of any other standard refrigerants.
Section \ref{sec5} is devoted to the rigorous presentation of the linear stability analysis that leads to a differential eigenvalue problem to be treated via the Chebyshev-$\tau$ method. In the last section, numerical results are discussed and compared with existing experimental data confirming that the model captures the main features of the physical phenomenon and frames the dryout as a consequence of interfacial instability.

\section{Problem framing}\label{sec2}

This section is intended to show a schematic representation of the physical setup discussed in the introduction for mathematical modelling purposes. 
Hence, let us consider a horizontal pipe of diameter $D$ where a two-phase annular flow develops. 
The flow involves two immiscible fluids: $\mathfrak{F}^{l}$, $\mathfrak{F}^{v}$, where the former is in contact with the cylinder walls and acts as refrigerant, the latter occupies the bulk and is generated by phase change (evaporation) of $\mathfrak{F}^{l}$. Moreover, the two fluids are separated by an interface, which is assumed to be smooth and whose profile changes as phase transition from $\mathfrak{F}^{l}$ to $\mathfrak{F}^{v}$ takes place.

\begin{figure}[!h]
    \centering
    \includegraphics[scale=0.5]{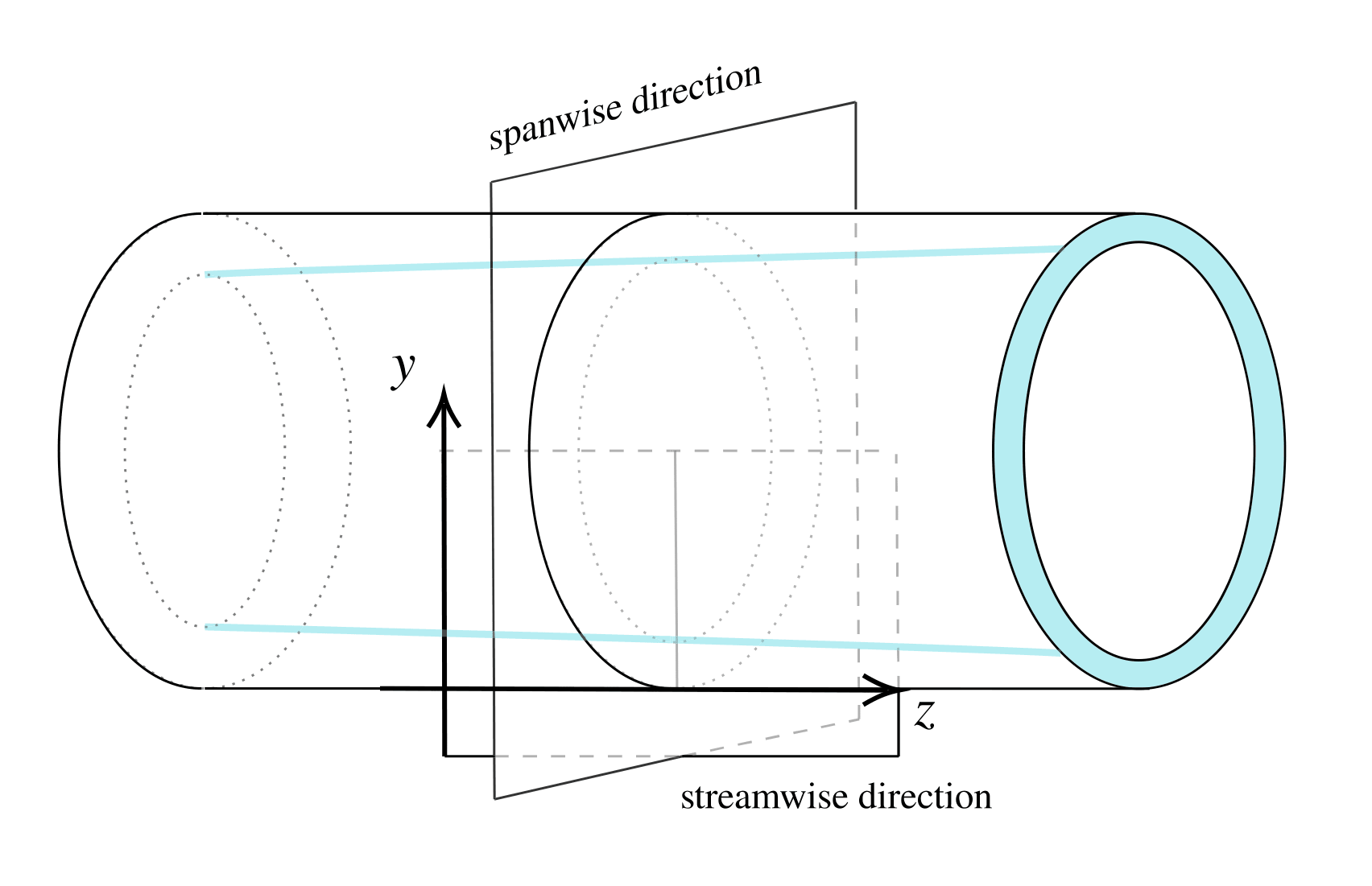}
    \caption{Sketch of distinct viewpoints for a cylindrical configuration. Streamwise point of view is adopted in this study.}
    \label{fig1}
\end{figure}

\begin{figure}[!h]
    \centering
    \includegraphics[scale=0.45]{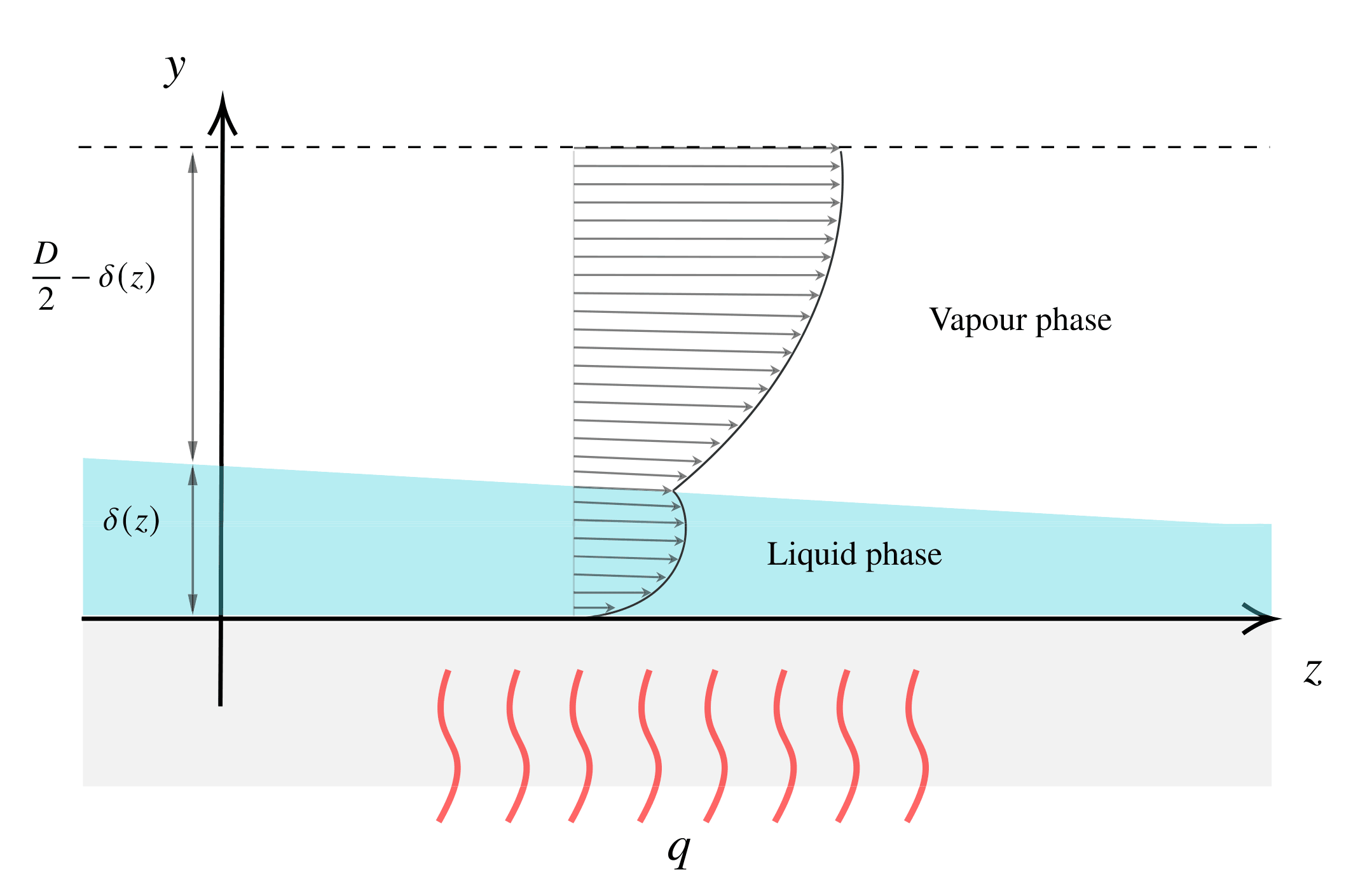}
    \caption{Sketch of physical configuration, where the liquid film is attached to the pipe wall, while the bulk is occupied by the vapour phase. The sloping interface between the two fluids is identified by $\delta(z)$. External heat is provided to the system via the heat flux $q$.}
    \label{fig2}
\end{figure}

To introduce the framework, let us notice that the physical setup can be viewed from two perspectives: streamwise and spanwise  (see Figure \ref{fig1}).
In the present investigation, for modelling purposes, a streamwise point of view is adopted. Moreover, given the cylindrical shape of the domain, it is assumed that the flow exhibits radial symmetry with respect to the central axis of the cylinder and  a longitudinal section of the pipe is considered. In this way, the domain reduces to a bidimensional horizontal layer of depth $D/2$, where the upper boundary is identified by the cylinder axis, while the lower boundary is the cylinder sidewall, subject to the presence of an external heat flux $q$. In this framework, as depicted in Figure \ref{fig2}, $\mathfrak{F}^{l}$ occupies the region closer to the lower boundary and the two fluids are separated by a smooth interface with decreasing profile. This profile identifies the phase transition that takes place during the cooling process, as the thickness of the liquid film decreases.

Let us now introduce a bidimensional Cartesian reference frame $Ozy$ with unit vectors $\{\textbf{k},\textbf{j}\}$, respectively, with $\textbf{j}$ pointing upward. 
The generic fluid $\mathfrak{F}^\alpha, (\alpha=l,v $ where $l$ stands for ``liquid" and $v$ for ``vapour"$)$ is characterized by dynamic viscosity $\mu^\alpha$, density $\rho^\alpha$, stress tensor $\bm{\sigma}^\alpha$ and the unknown fields are velocity and pressure $\textbf{v}^\alpha(z,y,t)=(u^\alpha,w^\alpha)$ and $p^\alpha=p^\alpha(z,y,t)$, respectively. Moreover, the interface separating the two superposed fluids is identified by the \textit{implicit equation} $S(\mathbf{x})=0$, where $\textbf{x}=(z,y)$. The implicit function theorem guarantees the existence of $y=\delta(z)$
such that 
\begin{equation}\label{def_S}
    S(z,\delta(z))=y-\delta(z)=0
\end{equation} 
Let us remark that  $y=\delta(z)$ is the equation that identifies the sloping interface separating the two fluids, see Figure \ref{fig2}.

\section{Interfacial conditions}\label{sec3}
  Here, the equations that guarantee conservation of mass, momentum and energy across the interface are derived from first-principles theory. 
    Indeed, the mathematical model arises from general conservation laws, which hold in each fluid phase and at the interface separating the two phases. 
    
    Let us define the quantity $\bm\psi^\alpha=[\psi^\alpha]_k, (k=1,\ldots, 5)$  within each fluid $\mathfrak{F}^\alpha, (\alpha=l,v)$.
    The general form of conservation laws for $\bm\psi^\alpha$ is:
    \begin{equation}\label{general_cons}
		\dfrac{\partial {\color{black}{\bm\psi^{\alpha}}}}{\partial t}+ \nabla \cdot\bm{f}^{\alpha} ={\color{black}{\mathbf{0}}}, 
	\end{equation}
	where $ \bm{f}^\alpha $ is the flux of $\bm\psi^\alpha$. 
    In continuum mechanics, by employing the eulerian point of view, Eq. \eqref{general_cons} is specialized by defining, for $i,j=1,2,3$,
	\begin{equation}\label{fluxes}
		[{f}^\alpha]_{ij}=[\psi^{\alpha}]_i[{v}^{\alpha}]_j+[{G}^\alpha]_{ij},
	\end{equation}
	and
	\begin{equation}\label{matrix1}
		{\color{black}{\bm\psi^{\alpha}}}=\left({\color{black}\begin{array}{c}
				\rho^{\alpha}\\[2mm]
				\rho^{\alpha} [{v}^{\alpha}]_i\\[2mm]
		\tfrac{1}{2}\rho^{\alpha} \lvert\textbf{v}^{\alpha}\rvert^2+\rho^{\alpha}\varepsilon^{\alpha}
		\end{array}}\right),\; {\color{black}{\textbf{G}}^{\alpha}}=\left({\color{black}{\begin{array}{c}
					0_j\\[2mm]
					-[\sigma^{\alpha}]_{ij}\\[2mm]
					-[\sigma^{\alpha}]_{jk}[{v}^{\alpha}]_k+[{Q}^{\alpha}]_j
		\end{array}}}\right)
	\end{equation}
    where $\rho^\alpha$ is the fluid density, $\textbf{v}^\alpha$ is the fluid velocity, $\bm\sigma^\alpha$ is the Cauchy stress tensor, $\varepsilon^\alpha$ is the internal energy and $\textbf{Q}^\alpha$ is the heat flux \cite{ruggeri2015rational}.
    The resulting balance equations, expressing the conservation of mass, momentum and energy, are:
	\begin{equation}\label{5}
		\small\begin{cases}
			\dfrac{\partial \rho^{\alpha}}{\partial t}+\nabla\cdot (\rho^{\alpha}\textbf{v}^{\alpha})=0		\\[2mm]
			\dfrac{\partial (\rho^{\alpha}\textbf{v}^{\alpha})}{\partial t}+\nabla\cdot (\rho^{\alpha} \textbf{v}^{\alpha}\otimes\textbf{v}^{\alpha}-\bm{\sigma}^{\alpha})=\bm{0},\\[2mm]
			\dfrac{\partial \left(\rho^{\alpha}\varepsilon^{\alpha}+\tfrac{1}{2}\rho (v^{\alpha})^2\right)}{\partial t}+\nabla\cdot \Bigl[\left(\rho^{\alpha}\varepsilon^{\alpha}+\tfrac{1}{2}\rho (v^{\alpha})^2\right)\textbf{v}^{\alpha}-\bm{\sigma}^{\alpha}\textbf{v}^{\alpha}+\textbf{Q}^{\alpha}\Bigr]= 0
		\end{cases} 
	\end{equation}

    \noindent
    and hold within each fluid $\mathfrak{F}^\alpha, (\alpha=l,v)$.
    
    Now, let us assume that densities $\bm\psi^\alpha$ are conserved at the interface, namely Eqs. \eqref{general_cons}-\eqref{matrix1} hold within a domain $\Omega$ astride the interface.     
    If $S(\textbf{x},t)=0$ is the equation that identifies the interface separating the two fluids, the following general conservation law at the interface holds \cite{hsieh1972effects}.
   
\begin{equation}\label{general_interface}
	\bm{f}^{l} \nabla S+\bm\psi^{l}\dfrac{\partial S}{\partial t}=\bm{f}^{v}\nabla S+\bm\psi^{v}\dfrac{\partial S}{\partial t}.
	\end{equation}
    By substituting Eqs. \eqref{fluxes}-\eqref{matrix1} into Eq. \eqref{general_interface},  and by considering the first component ($k=1$) of the resulting equation, the conservation of mass across the interface is obtained:
	\begin{equation}\label{cons_mass}
		\rho^{l}\left(\dfrac{\partial S}{\partial t}+\textbf{v}^{l}\cdot \nabla S\right)=\rho^{v}\left(\dfrac{\partial S}{\partial t}+\textbf{v}^{v}\cdot \nabla S\right).
	\end{equation}
    Index $k=2,3,4$ leads to the conservation of momentum across the interface:
	\begin{equation}\label{cons_momentum}
	\begin{split}
		\rho^{l}{\textbf{v}^{l}}\left(\dfrac{\partial S}{\partial t}+{\textbf{v}^{l}}\cdot\nabla S\right)&-{\bm{\sigma}^{l}} \nabla S=\rho^{v}{\textbf{v}^{v}}\left(\dfrac{\partial S}{\partial t}+{\textbf{v}^{v}}\cdot\nabla S\right)-{\bm{\sigma}^{v}} \nabla S.
	\end{split}
	\end{equation}
    At this stage, we should write the energy balance equation obtained at $k=5$. However, the resulting problem would be so complicated that a simplification step is needed.
    As anticipated in the introduction, we follow the idea of Hsieh \cite{hsieh1978interfacial}, which consists of replacing the conservation of energy with a density balance equation where the heat transfer is modelled as a production term of density of mass. Indeed, according to Hsieh \cite{hsieh1978interfacial}, transfer of mass across the interface represents a transformation from one phase to the other and therefore it is linked to latent heat associated to phase change. Consequently, thermal effects are modeled via mass transfer across the interface. Moreover, the position and modification of the interface corresponds to phase change associated with a certain latent heat. So,  thermal effects are strictly connected to the instantaneous position of the interface. Hence, the interfacial condition for energy balance is replaced by the following equation
	\begin{equation}\label{misterj}
		\rho^{l}\left(\dfrac{\partial S}{\partial t}+\textbf{v}^{l}\cdot \nabla S\right)=\dfrac{F(y)}{L},
	\end{equation}
    where $L$ is the latent heat required to convert liquid into vapour at constant temperature and $F(y)$ represents the net heat flux across the interface when phase change takes place \cite{hsieh1978interfacial,adham1995rayleigh}. Let us note that if $L\to\infty$, mass flux at the interface is absent as an infinite amount of heat would be required for a phase change to happen.

    Let us remark now that, since the actual physical setup involves microchannels, Stevin's contribution to pressure can be dropped compared to surface tension effects, therefore we assume $p-\rho g y \simeq p$, where $p$ is the \textit{absolute} pressure and $g$ the gravity. Therefore, the Cauchy stress tensor takes the following form:  
    \begin{equation}\label{extra_stress}
        \begin{split}
	[\bm{\sigma}^\alpha]_{hk}&=-p^\alpha \delta_{hk}+\mu^\alpha\left(\dfrac{\partial [v^\alpha]_h}{\partial x_k}+\dfrac{\partial [v^\alpha]_k}{\partial x_h}\right)	\\
		&=-p^\alpha \delta_{hk}+[\bar{\bm{\sigma}}^\alpha]_{hk}.
	\end{split}
    \end{equation}
    $\mu^\alpha$ being the dynamic viscosity of $\mathfrak{F}^\alpha$. 
    Therefore, Eq. \eqref{cons_momentum} becomes:
    \begin{equation}\label{cons_momentum2}
	\small \begin{split}
		\rho^{l}{\textbf{v}^{l}}\left(\dfrac{\partial S}{\partial t}+{\textbf{v}^{l}}\cdot\nabla S\right)&-{\bar{\bm{\sigma}}^{l}} \nabla S -(p^{v}-p^{l})\nabla S =\rho^{v}{\textbf{v}^{v}}\left(\dfrac{\partial S}{\partial t}+{\textbf{v}^{v}}\cdot\nabla S\right)-{\bar{\bm{\sigma}}^{v}} \nabla S.
	\end{split}
	\end{equation}
    Finally, let us assume that normal stresses due to pressure result in modification, near equilibrium, of the interface position and shape according to Young-Laplace equation:
    \begin{equation}\label{YL}
        \Delta p=\tau \left(\frac{1}{R_1}+\frac{1}{R_2}\right)
    \end{equation}
    where $\Delta p=p^{v}-p^{l}$ is the pressure difference at the interface, $\tau $ is the surface tension and $R_1, R_2$ are the principal radii of curvature.
    In conclusion, the interfacial conditions are the following:
    \begin{equation}\label{misterjakobson3}
		\small\begin{cases}
			\rho^{l}\left(\dfrac{\partial S}{\partial t}+\textbf{v}^{l}\cdot \nabla S\right)=\rho^{v}\left(\dfrac{\partial S}{\partial t}+\textbf{v}^{v}\cdot \nabla S\right),\\[4mm]
		\rho^{l}{\textbf{v}^{l}}\left(\dfrac{\partial S}{\partial t}+{\textbf{v}^{l}}\cdot\nabla S\right)-{\bar{\bm{\sigma}}^{l}} \nabla S -\tau \left(\dfrac{1}{R_1}+\dfrac{1}{R_2}\right)\nabla S =\rho^{v}{\textbf{v}^{v}}\left(\dfrac{\partial S}{\partial t}+{\textbf{v}^{v}}\cdot\nabla S\right)-{\bar{\bm{\sigma}}^{v}} \nabla S, \\[4mm]
			\rho^{l}\left(\dfrac{\partial S}{\partial t}+\textbf{v}^{l}\cdot \nabla S\right)=\dfrac{F(y)}{L},\\[4mm]
			{u}^{l}={u}^{v}
		\end{cases}
	\end{equation}
    where Eq. \eqref{misterjakobson3}$_4$ represents a regularity condition at the interface.

    Now, the governing equations that hold within each fluid $\mathfrak{F}^\alpha$ $(\alpha=l,v)$ are the incompressible Navier-Stokes equations:
    \begin{equation}
             \begin{cases}
		\nabla\cdot \textbf{v}^\alpha=0\\
        \dfrac{\partial \textbf{v}^\alpha}{\partial t}+\textbf{v}^\alpha\cdot \nabla \textbf{v}^\alpha=-\dfrac{1}{\rho^\alpha}\nabla p^\alpha+\dfrac{\mu^\alpha}{\rho^\alpha}\Delta \textbf{v}^\alpha
	\end{cases} 
        \end{equation}
        that can be derived from Eq. \eqref{5}$_{1-2}$, by virtue of the stress tensor in Eq. \eqref{extra_stress}.
        
        This system of governing equations is coupled with the following boundary conditions:
        \begin{equation}\label{mauricegentil}
\begin{aligned}
\mathbf{v}^{l} &= \bm{0} \quad \text{on } y=0 \\
\mathbf{v}^{v} &= V \textbf{k} \quad \text{on } y=\frac{D}{2}
\end{aligned}
        \end{equation}
        which models no-slip condition at the upper and lower boundaries, as if the velocity of the fictitious upper boundary is $\textbf{V}=V\textbf{k}$. 

\section{Equations of motion and dimensionless formulation}\label{sec4}

The nondimensional formulation of Navier-Stokes equations and interfacial conditions is obtained by setting the diameter $D$ of the tube as characteristic length, the ratio $G/\rho^{l}$  of mass flux over liquid density as the characteristic velocity, while the characteristic time is $(D\rho^{l})/G$ and the characteristic pressure is $P=(G)^2/\rho^{l}$. Hence, denoting by the asterisks the nondimensional quantities
\begin{equation}
    \textbf{x}=D\textbf{x}^*,\quad {\color{black}{\textbf{v}^\alpha=\dfrac{G}{\rho^{l}}(\textbf{v}^{\alpha})^*}}, \quad t=\dfrac{D\rho^{l}}{G}t^*,\quad p^\alpha=P({p^{\alpha}})^*,
\end{equation}
    the Navier-Stokes equations for both fluids become, omitting the asterisks for notational convenience,
    \begin{equation}
    \begin{cases}
         \dfrac{\partial\textbf{v}^{l}}{\partial t}+\textbf{v}^{l}\cdot \nabla \textbf{v}^{l}=-\nabla p^{l}+\dfrac{1}{\text{Re}^{l}}\Delta \textbf{v}^{l}\\[2mm]
        \dfrac{\partial\textbf{v}^{v}}{\partial t}+\textbf{v}^{v}\cdot \nabla \textbf{v}^{v}=-\widehat{\rho}\,\nabla p^{v}+\dfrac{1}{\text{Re}^{v}}\widehat{\rho}\Delta \textbf{v}^{v}
    \end{cases}
    \end{equation}
    Here $\widehat{\rho}=\tfrac{\rho^l}{\rho^v}$ and 
    \begin{equation}
       \text{Re}^\alpha=\dfrac{GD}{\mu^\alpha}
    \end{equation}
    which are the
    Reynolds numbers relative to the two fluids $\mathfrak{F}^\alpha, \alpha=l,v$.  They model the interplay between inertial and viscous forces, which act in different ways in the two fluids. Let us notice that the difference with a more standard definition of the Reynolds number is consequence of the choice of the characteristic velocity as $\frac{G}{\rho^{l}}$, working with the intensity of the mass flux rather than the mean velocity of the fluid.

    Therefore, by defining
    \begin{equation}
       \ell^\alpha=\begin{cases}
		1&\text{if}\; \alpha=l,\\
		\widehat{\rho}&\text{if}\; \alpha=v,
	\end{cases}
    \end{equation}
    the nondimensional Navier-Stokes equations can be written in short form:
    \begin{equation}
        \begin{cases}\label{pepposss}
          \nabla \cdot \textbf{v}^\alpha=0\\   \dfrac{\partial\textbf{v}^{\alpha}}{\partial t}+\textbf{v}^{\alpha}\cdot \nabla \textbf{v}^{\alpha}=-\ell^\alpha\nabla p^{\alpha}+\dfrac{\ell^\alpha}{\text{Re}^\alpha}\Delta \textbf{v}^{\alpha}            
        \end{cases}
    \end{equation}
    with boundary conditions Eq. \eqref{mauricegentil} assigned on $y=0,y=\frac{1}{2}$.
    The interfacial conditions, instead, omitting the asterisks, become
   
    \begin{equation}\label{misterjakobson4}
	\small \begin{cases}
		\dfrac{\partial S}{\partial t}+\textbf{v}^{l}\cdot \nabla S=\bar{\rho}\left(\dfrac{\partial S}{\partial t}+\textbf{v}^{v}\cdot \nabla S\right),\\[4mm]
	{\textbf{v}^{l}}\left(\dfrac{\partial S}{\partial t}+\textbf{v}^{l}\cdot \nabla S\right)-\dfrac{1}{\text{Re}^{l}}\bar{\bm{\sigma}}^{l} \nabla S=\bar{\rho}{\textbf{v}^{v}}\left(\dfrac{\partial S}{\partial t}+{\textbf{v}^{v}}\cdot\nabla S\right)\\
    \hspace{5cm}-\dfrac{1}{\text{Re}^{v}}\bar{\bm{\sigma}}^{v} \nabla S+\dfrac{1 }{\text{We}}\left(\dfrac{1}{R_1}+\dfrac{1}{R_2}\right)\nabla S,\\[4mm]
	\dfrac{\partial S}{\partial t}+\textbf{v}^{l}\cdot \nabla S=\dfrac{F(y)}{L},\\[4mm]
		{{u^{l}={u}^{v}	}}
	\end{cases}
\end{equation}
where $\bar{\rho}=\frac{\rho^v}{\rho^l}$
and the ratio $\frac{\mu^{l} G}{\rho^{l} D}$ has been chosen as characteristic quantity for the stress tensors $\bar{\bm{\sigma}}^\alpha$, while
\begin{equation}
     \text{We}=\dfrac{(G)^2D}{\tau \rho^{l}}
\end{equation}
is the Weber number.
This nondimensional number plays a pivotal role in analysing interfacial problems. Indeed, it is defined as the ratio between inertial forces 
and forces due to surface tension.

The Navier-Stokes equations \eqref{pepposss} admit a steady solution denoted by $m_b^\alpha=(\textbf{v}_b^\alpha,p_b^\alpha)$, where $\textbf{v}^\alpha_b(z,y)=(u^\alpha_b,w^\alpha_b)$ and the interface at the steady state is identified by Eq. \eqref{def_S}. Let us remark that the velocity profile of liquid and vapour phases satisfy Eq. \eqref{misterjakobson4}$_4$.   Therefore, at the interface  $y=\delta(z)$, $\textbf{v}^\alpha_b(z,\delta(z))=\textbf{V}_b(z,\delta(z))$ where  $\textbf{V}_b=(U_b,W_b)$. Nevertheless, their derivatives are different in general, therefore derivatives of $\textbf{v}_b^\alpha$ at the interface are denoted by e.g. $(\textbf{v}^\alpha_b)_y=(\textbf{V}^\alpha_b)_y$.

At this stage of the study, we do not specify the basic velocity profile of the two fluids. This choice is motivated by the desire of keeping the analysis as general and broadly applicable as possible. Therefore, we proceed to perform a linear stability analysis of this generic profile in order to determine a general form of the perturbation equations.
However, numerical investigation is performed by selecting base velocity profiles $\textbf{v}^{\alpha}_b$ consistent with annular flow.

\subsection{The dryout instability factor}

In this paragraph, the main steps to determine the dimensionless dryout instability factor defined in \cite{cantini2025inception} are outlined. The reader is referred to this paper for a detailed derivation.
Given Eq. \eqref{def_S}, it follows that
\begin{equation}\label{mg}
    \frac{\partial S}{\partial z} = -\frac{\text{d}\delta}{\text{d}z}
\end{equation}
For the sake of convenience, the function $\delta(z)$ is expressed as a composite function
$\delta=\delta(\alpha(x(z)))$ so that the chain rule can be applied:  
    \begin{equation}
    \label{eq:derivativeofdelta}
        \frac{\text{d}\delta}{\text{d}z}=\frac{\text{d}\delta}{\text{d}\alpha}\frac{\text{d}\alpha}{\text{d}x}\frac{\text{d}x}{\text{d}z}
    \end{equation} 
where $\alpha$ is the void fraction and $x$ the vapour quality. The derivatives in Eq. \eqref{eq:derivativeofdelta} can be calculated as in \cite{cantini2025inception} and are:
\begin{equation}\label{eqs}
    \begin{split}
        & \frac{\text{d}\delta}{\text{d}\alpha} = -\frac{D}{4\sqrt{\alpha}}\\
        & \frac{\text{d}\alpha}{\text{d}x} = \frac{\alpha\left(1-\alpha\right)}{x\left(1-x\right)}\\
        & \frac{\text{d}x}{\text{d}z}=\frac{4q}{h_{gl}GD}
    \end{split}
\end{equation}
where $h_{gl}$ is the relative enthalpy between liquid and gaseous phases.
Substituting Eqs. \eqref{mg}-\eqref{eqs} in the last term of Eq. \eqref{misterjakobson4}$_2$ projected on the $z$ axis and being $
R_1= \frac{D}{2}\sqrt{\alpha}$, $R_2=\infty$, the following expression is obtained:
    \begin{equation}
        \frac{1}{\text{We}}\left(\frac{1}{R_1}+\frac{1}{R_2}\right)\frac{\partial S}{\partial z} = \frac{2\left(1-\alpha\right)}{x\left(1-x\right)}\frac{\text{Bo}}{\text{We}}
        \label{eq:dryfactor}
    \end{equation}
The Boiling number  
\[ \text{Bo} =\frac{q}{h_{gl}G} \]
is defined as the ratio between the rate of phase change over the total mass flow rate. 
while the ratio
\[ \frac{\text{Bo}}{\text{We}}=\frac{\rho^{l}\tau q}{G^3h_{gl}D} \]
exhibits an inverse proportionality to the mass flux. This mathematical form is consistent with the experimentally--derived $\delta^+$ behaviour \cite{cantini2025inception}. 
In particular, it is shown that the dryout vapour quality $x_{dry}$ is correlated with the right-hand-side of Eq. \eqref{eq:dryfactor}, defined as \textit{dryout instability factor}
\begin{equation}\label{cantinis}
  I_{\delta^+}:= \frac{2\left(1-\alpha\right)}{x\left(1-x\right)}\frac{\text{Bo}}{\text{We}}
\end{equation} 
for the $\delta^+$ regime. Let us remark that the importance of this number lies also in the geometrical configuration that has been considered. In millichannels, the surface tension plays a pivotal role, whereas if a larger diameter tube were considered, the surface tension would be small and $I_{\delta^+}$ would be negligible too.

\section{CO\textsubscript{2} as the reference fluid for the \texorpdfstring{$\delta^+$}{delta+} regime}\label{tabelle}

This section aims to give a detailed discussion on the reason why the object of the present study focuses on two-phase CO\textsubscript{2} annular flow in the $\delta^+$ regime. As already underlined in the introduction, carbon dioxide uniquely accesses, under practically relevant evaporator conditions, the thermodynamic state required for the onset of the $\delta^+$ regime. It is now essential to clarify why CO\textsubscript{2} can exhibit a different trend compared to other refrigerants.

A comparison with conventional refrigerants (e.g. R-12, R-134a, R-290) shows that, at the same evaporator conditions, unlike other parameters, liquid-to-vapour density ratio and reduced pressure vary significantly.
In particular, CO\textsubscript{2} at typical refrigeration temperatures operates at much higher reduced pressure and exhibits a much smaller density ratio (i.e. more similar liquid and vapour densities) than conventional refrigerants. This is the key discriminating feature.
Table \ref{firstable} summarizes the comparison at identical standard conditions. The key point is that We, Re, Bd and Bo remain essentially comparable in magnitude across fluids, while $\rho^l/\rho^v$ and $p_r$ differ significantly.

\begin{table}[htbp]
\centering
\begin{tabular}{lcccccc}
\hline
\textbf{Fluid} & \textbf{We} & \textbf{Re} & \textbf{Bd} & \textbf{Bo} & $\boldsymbol{\rho^l/\rho^v}$ & $\boldsymbol{p_r}$ \\
 & $(-)$ & $(-)$ & $(-)$ & $(-)$ & $(-)$ & $(-)$ \\
\hline
CO\textsubscript{2} & 194,189 & 9,311 & 1.263 & 0.0154 & \textbf{16.6} & \textbf{0.359} \\
R-12             & 71,969  & 4,157 & 1.013 & 0.0261 & 132.5 & 0.056 \\
R-134a           & 79,032  & 3,697 & 0.965 & 0.0199 & 162.0 & 0.050 \\
R-290            & 217,579 & 8,175 & 0.440 & 0.0106 & 84.3  & 0.056 \\
\hline \\
\end{tabular}
\caption{Dimensionless parameters for all refrigerants at standard conditions ($T_{\text{sat}}=-15^\circ$C, $G=1200$ kg\,m$^{-2}$\,s$^{-1}$, $D=1$ mm, $q=5$ kW\,m$^{-2}$), data collected from \cite{nist_webbook}. CO\textsubscript{2} is a clear outlier in terms of liquid-to-vapour density ratio and reduced pressure, indicating a near-critical operating state with much more similar phase densities.}
\label{firstable}
\end{table}


This shows that CO\textsubscript{2} is the only fluid among standard refrigerants that realizes, at useful evaporator temperatures, the low density-ratio condition required for $\delta^+$ behaviour.  The small density ratio achieved by CO\textsubscript{2} reduces phase slip and weakens interfacial shear. This suppresses droplet entrainment, helps preserve a coherent annular film, and allows increasing mass flux to enhance convective evaporation in the film without immediately disrupting it. Under these conditions, dryout can be delayed as $G$ increases, which is precisely the signature of the $\delta^+$ regime.

The same conclusion can be drawn from Table~\ref{tab:co2_standard_percent}, where the percentage deviation relative to CO\textsubscript{2} is reported. In fact, variations in We, Re, Bd and Bo are significant but remain within factors of order unity. By contrast, the density ratio differs significantly, which yields different qualitative annular-flow dynamics.

\begin{table}[htbp]
\centering
\begin{tabular}{lcccccc}
\hline
\textbf{Fluid} & \textbf{We} & \textbf{Re} & \textbf{Bd} & \textbf{Bo} & $\boldsymbol{\rho_l/\rho_g}$ & $\boldsymbol{p_r}$ \\
 & (\%) & (\%) & (\%) & (\%) & (\%) & (\%) \\
\hline
CO\textsubscript{2} & --    & --    & --    & --    & --  & -- \\
R-12             & $-$62.9 & $-$55.4 & $-$19.8 & $+$69.5 & $+$698 & $-$84.4 \\
R-134a           & $-$59.3 & $-$60.3 & $-$23.6 & $+$29.2 & $+$876 & $-$86.1 \\
R-290            & $+12.0$ & $-12.2$ & $-65.2$ & $-31.2$ & $+408$ & $-84.4$ \\
\hline\\
\end{tabular}
\caption{Percentage differences from CO\textsubscript{2} at the same standard conditions. The key discrepancy is the liquid-to-vapour density ratio, which differs by $+$408\% to $+$876\%, far exceeding the variation of the classical dimensionless groups.}
\label{tab:co2_standard_percent}
\end{table}


We now want to underline that the observed difference is not due to operating conditions. In fact, let us focus on R-290 so as to reproduce the  CO\textsubscript{2} reference case by adjusting the operating conditions (channel diameter, mass flux, and wall heat flux) so as to match the standard dimensionless groups as closely as possible. This is a stringent test of classical similitude arguments:
if We, Re, Bd, Bo numbers were sufficient to reproduce similar values for the density ratio, then the R-290 case should reproduce the same $\delta^+$ behaviour as CO\textsubscript{2}.

Table~\ref{tab:co2_optimization_attempts} shows that this is not the case. Even when We, Re, Bd, and Bo are matched with excellent accuracy, the density ratio remains much larger than in CO\textsubscript{2}. This remains true both at the same saturation temperature and at elevated saturation temperature.

\begin{table}[htbp]
\centering
\footnotesize
\begin{tabular}{lcccccccc}
\hline
\textbf{Fluid} & $\textbf{D}$ & $\textbf{G}$ & $q$ & $\textbf{We}$ & $\textbf{Re}$ & $\textbf{Bd}$ & $\textbf{Bo}$ & $\boldsymbol{\rho^l/\rho^g}$ \\
 & (mm) & (kg/m$^2$s) & (kW/m$^2$) & $(-)$ & $(-)$ & $(-)$ & $(-)$ & $(-)$ \\
\hline
CO\textsubscript{2} ($-15^\circ$C) & 1.000 & 1200 & 5.0 & 194 & 9  & 1.263 & 0.0154 & \textbf{16.6} \\
R-290 ($-15^\circ$C)               & 1.694 & 871  & 5.3 & 194 & 10 & 1.263 & 0.0154 & 84.3 \\
R-290 ($+15^\circ$C)               & 1.469 & 744  & 4.0 & 194 & 10 & 1.263 & 0.0154 & 32.1 \\
\hline\\
\end{tabular}
\caption{Optimization attempts to match CO\textsubscript{2} dimensionless parameters using R-290 by adjusting $D$, $G$, and $q$. Although We, Re, Bd, and Bo can be matched closely, the liquid-to-vapour density ratio remains substantially higher than for CO\textsubscript{2}.}
\label{tab:co2_optimization_attempts}
\end{table}

This demonstrates two key points:
\begin{enumerate}
    \item matching the classical dimensionless groups is possible by tuning operating and geometric parameters;
    \item matching the density ratio is not possible at practical evaporator temperatures.
\end{enumerate}

This shows that the density ratio is a key parameter for the $\delta^+$ regime. 
In conclusion, to achieve density ratios comparable to CO\textsubscript{2}, conventional refrigerants would need to operate much closer to their critical points, i.e. at substantially higher saturation temperatures. For fluids such as R-290, this would require evaporator temperatures far above the practical range used in refrigeration and air-conditioning applications.
Such conditions are thermodynamically irrelevant for cooling systems: an evaporator operating at very high saturation temperature provides little or no useful cooling relative to typical heat-sink temperatures. Therefore, even if a $\delta^+$ behaviour were theoretically possible for another refrigerant, it would occur outside the range where the evaporator is used as a refrigeration device.
This is the key reason why CO\textsubscript{2} is the natural and practically relevant fluid for the present study. Its low critical temperature allows near-critical operation (and hence low density ratio) \emph{within} the temperature range where evaporators are actually employed for cooling.

The rationale for using CO\textsubscript{2} in this work can therefore be summarized as follows:
\begin{enumerate}
    \item The $\delta^+$ regime is primarily controlled by thermodynamic state, especially phase-density similarity, rather than by fluid chemistry.
    \item CO\textsubscript{2} uniquely provides low liquid-to-vapour density ratio at practical evaporator temperatures because of its low critical temperature.
    \item Conventional refrigerants can match standard dimensionless groups but not the density ratio under useful refrigeration conditions.
    \item The present interfacial-stability-based model is physically consistent precisely in the low-density-ratio annular-flow conditions that CO\textsubscript{2} can realize.
\end{enumerate}

For these reasons, CO\textsubscript{2} is not simply one refrigerant among many in the present study: it is the physically and technologically relevant working fluid for investigating the $\delta^+$ dryout regime in refrigeration-scale evaporators.

\section{Linear analysis}\label{sec5}

In this section, the linear stability analysis is presented upon the introduction of small amplitude perturbations on the unknown fields. Following \cite{adham1995rayleigh}, also the interface is perturbed by a periodic function of infinitesimally small amplitude. This will lead to a couple of forth-order differential equations that require four conditions at the interface and four boundary conditions.

\subsection{Linearised governing equations for perturbations}

Let us introduce a perturbation on the steady solution $m_b^\alpha$, defined as follows
\begin{equation}\label{perturb0}
    \textbf{v}^\alpha(z,y,t) = \textbf{v}^\alpha_b(z,y) + \widehat{\textbf{v}}^\alpha(z,y,t), \quad p^\alpha(z,y,t) = p^\alpha_b(z,y) + \widehat{p}^\alpha(z,y,t)
\end{equation}
	where $\widehat{\textbf{v}}^\alpha(z,y,t)=(\widehat{u}^\alpha,\widehat{w}^\alpha)$. 

    Substituting Eq. \eqref{perturb0} into Eq. \eqref{pepposss}, the resulting system is autonomous, therefore perturbation fields take the following form:
\begin{equation}\label{perturb}
    \widehat{\textbf{v}}^\alpha(z,y,t)=\widetilde{\textbf{v}}^\alpha(y)e^{ikz-nt}, \quad  \widehat{p}^\alpha(z,y,t)=\widetilde{p}^\alpha(y)e^{ikz-nt} 
\end{equation}
	where $\widetilde{\textbf{v}}^\alpha(y)=(\widetilde{u}^\alpha,\widetilde{w}^\alpha )$, $k$ is the wavenumber and $n=n_r+in_i$ with $n_r$ being the growth rate.

By virtue of Eqs. \eqref{perturb0}-\eqref{perturb}, the continuity equation Eq. \eqref{pepposss}$_1$, becomes:
	\begin{equation}\label{MF}
		\wt{u}^\alpha(y)=-\dfrac{\dot{\wt{w}}^\alpha(y)}{ik}
	\end{equation}
    which is a useful relation between the first and second components of velocity perturbation and will allow us to reduce the number of unknown fields.
Moreover, by virtue of \eqref{perturb0}, the momentum equation Eq. \eqref{pepposss}$_2$, dropping nonlinear terms, becomes
\begin{equation}\label{josephline}
    \begin{split}
	\wh{{u}}^\alpha_{t}+{u}_b^\alpha\wh{{u}}^\alpha_z+{w}_b^\alpha\wh{{u}}^\alpha_y +\wh{u}^\alpha({u}_b^\alpha)_z+\wh{w}^\alpha({u}_b^\alpha)_y &=-\ell^\alpha\wh{p}^\alpha_z+\dfrac{1}{\text{Re}^\alpha}\ell^\alpha\Delta \wh{u}^\alpha,\\
	\wh{{w}}^\alpha_{t}+{u}_b^\alpha\wh{{w}}^\alpha_z+{w}_b^\alpha\wh{{w}}^\alpha_y +\wh{u}^\alpha({w}_b^\alpha)_z+\wh{w}^\alpha({w}_b^\alpha)_y  &=-\ell^\alpha\wh{p}^\alpha_y+\dfrac{1}{\text{Re}^\alpha}\ell^\alpha\Delta \wh{w}^\alpha.
\end{split} 
\end{equation} 
Now, by deriving \eqref{josephline}$_1$ with respect to $y$ and \eqref{josephline}$_2$ with respect to $z$ and, by summing the resulting equations, by virtue of the incompressibility constraint, we get
\begin{equation}\label{NS_perturbed}
    \begin{split}
	&\wh{{u}}^\alpha_{ty}+{u}_b^\alpha\wh{{u}}^\alpha_{zy}+{w}_b^\alpha\wh{{u}}^\alpha_{yy}  +\wh{u}^\alpha({u}_b^\alpha)_{zy}+\wh{w}^\alpha({u}_b^\alpha)_{yy}-\dfrac{1}{\text{Re}^\alpha}\ell^\alpha (\wh{u}^\alpha_{zzy}+\wh{u}^\alpha_{yyy})\\
	&\qquad=\wh{{w}}^\alpha_{tz}+{u}_b^\alpha\wh{{w}}^\alpha_{zz}+{w}_b^\alpha\wh{{w}}^\alpha_{yz} +\wh{u}^\alpha({w}_b^\alpha)_{zz}+\wh{w}^\alpha({w}_b^\alpha)_{yz}-\dfrac{1}{\text{Re}^\alpha}\ell^\alpha (\wh{w}^\alpha_{xxx}+\wh{w}^\alpha_{yyx}).
\end{split}
\end{equation}

By virtue of Eqs. \eqref{perturb}-\eqref{MF}, Eq. \eqref{NS_perturbed} multiplied by $ik$, becomes:
\begin{equation}
    \begin{split}
&{\color{black}{n}}\left[D^2-k^2\right]{\wt{w}}^\alpha+{\color{black}{\dfrac{1}{\text{Re}^\alpha}\ell^\alpha}}\left[D^2-k^2\right]^2{\wt{w}}^\alpha-{\color{black}{ik}}\left((w^\alpha_{b})_{yz}-(u^\alpha_b)_{yy}\right){\wt{w}}^\alpha\\
&\qquad-{\color{black}{w_b^\alpha}}\left[D^2-k^2\right]{\color{black}{D}}{\wt{w}}^\alpha+\left((w_b^\alpha)_{zz}-(u_b^\alpha)_{zy}\right){\color{black}{D}}{\wt{w}}^\alpha-{\color{black}{iku_b^\alpha}}\left[D^2-k^2\right]{\wt{w}}^\alpha=0.
\end{split}
\end{equation}
with $D=\tfrac{d}{dy}$. Hence, the perturbed linear governing equations are, in compact form: 
\begin{equation}\label{NS-linear}
    \left(\begin{array}{cc}
		\mathcal{X}	& \mathcal{Y}
	\end{array}\right)\left(\begin{array}{c}
		\wt{w}^{l}\\
		\wt{w}^{v}
	\end{array}\right)=n\left(\begin{array}{cc}
		\mathcal{W}	& \mathcal{W}
	\end{array}\right)\left(\begin{array}{c}
		\wt{w}^{l}\\
		\wt{w}^{v}
	\end{array}\right) 
\end{equation}
where
\begin{equation}
    \begin{split}
\mathcal{X}&={\color{black}{\dfrac{1}{\text{Re}^{l}}}}\left[D^2-k^2\right]^2-{\color{black}{ik}}\left((w^{l}_{b})_{yz}-(u^{l}_b)_{yy}\right)-{\color{black}{w_b^{l}}}\left[D^3-k^2D\right]	\\
&\qquad\qquad\qquad +\left((w_b^{l})_{zz}-(u_b^{l})_{zy}\right){\color{black}{D}}-{\color{black}{iku_b^{l}}}\left[D^2-k^2\right]\\
\mathcal{Y}&={\color{black}{\dfrac{1}{\text{Re}^{v}}\widehat{\rho}}}\left[D^2-k^2\right]^2-{\color{black}{ik}}\left((w^{v}_{b})_{yz}-(u^{v}_b)_{yy}\right)-{\color{black}{w_b^{v}}}\left[D^3-k^2D\right]	\\
&\qquad\qquad\qquad +\left((w_b^{v})_{zz}-(u_b^{v})_{zy}\right){\color{black}{D}}-{\color{black}{iku_b^{v}}}\left[D^2-k^2\right]\\
\mathcal{W}&=-\left[D^2-k^2\right]
\end{split}
\end{equation}

\subsection{Linearised interfacial conditions for perturbations}

Let us introduce a perturbation $\zeta=\zeta(z,t)$ on the interface so that Eq. \eqref{def_S} is replaced by:    	
	\begin{equation}\label{perturb_interface}
		S(\textbf{x},t)=y-\delta(z)-\zeta(z,t),
	\end{equation}
    and the position of the perturbed interface is identified by $S(\textbf{x},t)=0$. Hence, the interfacial condition \eqref{misterjakobson4}$_3$ becomes:
	\begin{equation}\label{mister}
		\rho^{l}\left(\dfrac{\partial S}{\partial t}+\textbf{v}^{l}\cdot \nabla S\right)=\dfrac{F(\delta+\zeta)}{L}.
	\end{equation}
 It is possible to recover the following expression for $F(\delta+\zeta)$:
    \begin{equation}\label{alphazeta}
        \frac{F(\delta+\zeta)}{L}=\beta\zeta
    \end{equation}
    where $\beta=\frac{H}{L}\left(\tfrac{1}{D/2-\delta(z)}+\tfrac{1}{\delta(z)}\right)$. 
By virtue of Eqs. \eqref{perturb0}-\eqref{perturb} and by assuming $\zeta(z,t)=\xi e^{ikz-nt}$, Eqs.\eqref{mister}-\eqref{alphazeta} become, dropping nonlinear terms:
	\begin{equation}
	    -n{\color{black}{\xi}} -ik{\color{black}{\xi}} U_b^{l} +(ik)^{-1}\delta'\dot{\wt{w}}^{l} +\wt{w}^{l} =\beta {\color{black}{\xi}}.
	\end{equation}
	Setting $\gamma=-(ik)^{-1}$, we get: 
	\begin{equation}\label{csi}
	  {\color{black}{\xi}}=\dfrac{-\gamma \delta'\dot{\wt{w}}^{l}+\wt{w}^{l}}{\beta+n+ikU_b^{l}}.   
	\end{equation}
Moreover, by virtue of \eqref{misterjakobson4}$_1$, following the same reasoning, we have
\begin{equation}
    {\color{black}{\xi}}=\dfrac{\bar{\rho}(-\gamma \delta'\dot{\wt{w}}^{v}+\wt{w}^{v})}{\beta+\bar{\rho}(n+ikU_b^{v})}
\end{equation}
and the following equation is immediately obtained:
\begin{equation}\label{eq1}
    \dfrac{-\gamma \delta'\dot{\wt{w}}^{l}+\wt{w}^{l}}{\beta+n+ikU_b}=\dfrac{\bar{\rho}(-\gamma \delta'\dot{\wt{w}}^{v}+\wt{w}^{v})}{\beta+\bar{\rho}(n+ikU_b)}. 
\end{equation}
Let us define $ {\color{black}{\lambda=\beta+ik U_b}} $ and $ {\color{black}{\bar{\lambda}=\bar{\rho}^{-1}(\beta+\bar{\rho}ikU_b)}} $, so that \eqref{eq1} can be written as follows
\begin{equation}
    \begin{split}
	-\bar{\lambda}\gamma \delta'\dot{\wt{w}}^{l}+\bar{\lambda}\wt{w}^{l}+\lambda	\gamma \delta'\dot{\wt{w}}^{v}-\lambda\wt{w}^{v} = n\left[-\gamma \delta'\dot{\wt{w}}^{v}+\wt{w}^{v}+\gamma \delta'\dot{\wt{w}}^{l}-\wt{w}^{l}\right].
\end{split}	 
\end{equation}
In matrix compact form, it becomes:
\begin{equation}\label{ICfirst}
   \left(\begin{array}{cc}
\mathcal{A}^1	& \mathcal{B}^1
\end{array}\right)\left(\begin{array}{c}
\wt{w}^{l}\\
\wt{w}^{v}
\end{array}\right)=n\left(\begin{array}{cc}
\mathcal{C}^1	& \mathcal{D}^1
\end{array}\right)\left(\begin{array}{c}
\wt{w}^{l}\\
\wt{w}^{v}
\end{array}\right) 
\end{equation}
where
\begin{equation}\label{flora1_new}
   \begin{split}
\mathcal{A}^1&=
	\bar\lambda-\bar{\lambda}\gamma \delta' D \\
\mathcal{B}^1&= -\lambda+\lambda\gamma \delta'D\\
\mathcal{C}^1&=-1+\gamma\delta'D\\	
\mathcal{D}^1&=1-\gamma\delta'D
\end{split} 
\end{equation}  
Therefore, we provide the final perturbed interfacial condition relative to momentum, where nonlinear terms have been dropped. In particular, the equation for the first component of \eqref{misterjakobson4}$_2$ is
\begin{equation}\label{maurizio}
   \left(\begin{array}{cc}
\mathcal{A}^2	& \mathcal{B}^2
\end{array}\right)\left(\begin{array}{c}
\wt{w}^{l}\\
\wt{w}^{v}
\end{array}\right)=n\left(\begin{array}{cc}
\mathcal{C}^2	& \mathcal{D}^2
\end{array}\right)\left(\begin{array}{c}
\wt{w}^{l}\\
\wt{w}^{v}
\end{array}\right) 
\end{equation}
where
\begin{equation}\label{flora1}
   \begin{split}
\mathcal{A}^2&=\lambda \left([C_1+B]+[E_1-\gamma\delta' B]D+ F_1 D^2\right) \\
\mathcal{B}^2&=-\lambda \left(C_2+ E_2D+ F_2D^2\right)\\
\mathcal{C}^2&=-C_1-A+[-E_1+\gamma \delta' A]D-F_1D^2\\	
\mathcal{D}^2&=C_2+E_2D+F_2D^2
\end{split} 
\end{equation}  
and
\begin{equation}\label{ABC1}
 \small\begin{split}
     {\color{black}{A}}&=U_b(\bar{\rho}-1),\\
     {\color{black}{B}}&=-ik(U_b)^2+ik\bar{\rho}(U_b)^2+\dfrac{1}{\text{Re}^{l}}[2(U_b^{l})_z]ik-\dfrac{1}{\text{Re}^{v}}[2(U_b^{v})_z]ik-\dfrac{1}{\text{We}}k^2 \delta'+\dfrac{1}{\text{We}}\left(\dfrac{1}{R_1}+\delta''\right)ik,\\
     {\color{black}{C_1}}&=U_b-\dfrac{1}{\text{Re}^{l}}ik,\quad {\color{black}{C_2}}=\bar{\rho}U_b-\dfrac{1}{\text{Re}^{v}}ik,\\
     {\color{black}{E_1}}&=-2\gamma U_b\delta'+W_b\gamma+\dfrac{2}{\text{Re}^{l}}ik\gamma\delta',\quad {\color{black}{E_2}}=-2\gamma U_b\bar{\rho}\delta'+W_b\gamma\bar{\rho}+\dfrac{2}{\text{Re}^{v}}ik\gamma\delta',\\
     {\color{black}{F_1}}&=-\dfrac{1}{\text{Re}^{l}}\gamma,\quad {\color{black}{F_2}}=-\dfrac{1}{\text{Re}^{v}}\gamma.
 \end{split}   
\end{equation}
While, the equation for the second component of \eqref{misterjakobson4}$_2$ is
\begin{equation}\label{maurizio2}
   \left(\begin{array}{cc}
		{\mathcal{A}}^3	& {\mathcal{B}}^3
	\end{array}\right)\left(\begin{array}{c}
		\wt{w}^{l}\\
		\wt{w}^{v}
	\end{array}\right)=n\left(\begin{array}{cc}
		{\mathcal{C}}^3	& {\mathcal{D}}^3
	\end{array}\right)\left(\begin{array}{c}
		\wt{w}^{l}\\
		\wt{w}^{v}
	\end{array}\right)  
\end{equation} 
where ${\mathcal{A}}^3,{\mathcal{B}}^3,{\mathcal{C}}^3,{\mathcal{D}}^3$ are defined as in Eq. \eqref{flora1} but Eq. \eqref{ABC1} is replaced  by
\begin{equation}\label{ABC2}
  \small  \begin{split}
    {\color{black}{ {A}}}&=W_b(\bar{\rho}-1),  \\
    {\color{black}{ {B}}}&=-ikW_bU_b+\dfrac{1}{\text{Re}^{l}}[(U_b^{l})_y+(W_b^{l})_z]ik+\dfrac{1}{\text{We}}k^2+ik\bar{\rho}U_bW_b-\dfrac{1}{\text{Re}^{v}}[(U_b^{v})_y+(W_b^{v})_z]ik,\\
    {\color{black}{ {C}_1}}&=2W_b-\delta'U_b+\dfrac{1}{\text{Re}^{l}}\delta'ik,\quad {\color{black}{ {C}_2}}=2\bar{\rho}W_b-\bar{\rho}U_b\delta'+\dfrac{1}{\text{Re}^{v}}ik\delta',\\
    {\color{black}{ {E}_1}}&=-\gamma W_b\delta'-\dfrac{2}{\text{Re}^{l}},\quad {\color{black}{ {E}_2}}=-\gamma W_b\bar{\rho}\delta'-\dfrac{2}{\text{Re}^{v}},\\
    {\color{black}{ {F}_1}}&=\dfrac{1}{\text{Re}^{l}}\gamma\delta',\quad {\color{black}{ {F}_2}}=\dfrac{1}{\text{Re}^{v}}\gamma\delta'.
    \end{split}
\end{equation}
Let us remark that the last interfacial condition Eq. \eqref{misterjakobson4}$_4$ is simplified by virtue of Eq. \eqref{MF} and becomes:
\begin{equation}\label{maurizio3}
   \left(\begin{array}{cc}
		D	& -D
	\end{array}\right)\left(\begin{array}{c}
		\wt{w}^{l}\\
		\wt{w}^{v}
	\end{array}\right)=n\left(\begin{array}{cc}
		0	& 0
	\end{array}\right)\left(\begin{array}{c}
		\wt{w}^{l}\\
		\wt{w}^{v}
	\end{array}\right)  
\end{equation}

\section{Numerical method and results}\label{sec6}

This section is intended to comment on the numerical results obtained by solving the differential eigenvalue problem made of Eqs. \eqref{NS-linear}, \eqref{ICfirst}, \eqref{maurizio},  \eqref{maurizio2}, \eqref{maurizio3} together with boundary conditions. In a first instance, the base flow is assigned to liquid and vapour phases. Then, the numerical method is illustrated and validated against the Orr-Sommerfeld problem studied in \cite{DongarraStraughWalker1996}.
Afterwards, the main steps of the numerical procedure are outlined and
the comparison between numerical and experimental data is discussed.

\subsection{The base flow}\label{subsection_baseflow}
Let us define explicitly the base velocity in the liquid and vapour phases. The following hypotheses have been enforced:
\begin{enumerate}
    \item The horizontal velocity in the vapour core has a flat profile along the radius
    \item The horizontal velocity in the liquid film is linear: 0 at the wall and equal to the vapour velocity at the interface
\end{enumerate}
These profiles are consistent with experimental evidence. In particular, the Reynolds number in the vapour core is sufficiently high that a uniform velocity profile is a reasonable approximation. Whereas, the liquid film is very thin, therefore, as a first approximation, the velocity can be assumed to vary linearly over the film thickness.


Under these hypotheses, the base velocities $\textbf{v}^\alpha_b=(u_b^\alpha,w_b^\alpha)$ for $\alpha=l,v$ can be expressed in terms of the liquid film thickness $\delta$ and the vapour quality $x$.
The horizontal component of the vapour phase velocity is obtained by definition:
\begin{equation}
    u_b^{v} =
    \frac{G}{\rho^v}\frac{x}{\alpha}
\end{equation}
which, in dimensionless form, becomes:
\begin{equation}\label{eq:uv}
    u_b^{v*}=u_b^{v}\frac{\rho^l}{G}=\frac{\rho^l}{\rho^v}\frac{x}{\alpha}
\end{equation}
The horizontal velocity profile in the liquid film is linear, as mentioned in the previous hypothesis n. 2,
\begin{equation}\label{eq:ul}
    u_b^{l*}(r) = u_b^{v*}\frac{\frac{D}{2}-r}{\delta}, \qquad r\in\left[\frac{D}{2}-\delta,\frac{D}{2}\right]
\end{equation}
being $\delta=D\delta^*$.
From the horizontal profile, by virtue of the continuity equation and the condition stating that the velocity has to be parallel to the interface:
\begin{equation}\label{eq:62}
    \frac{W_b(z,\delta(z))}{U_b(z,\delta(z))}=\frac{\text{d} \delta}{\text{d}z}
\end{equation}
the expressions for the vertical components can be determined. Once the velocities profiles have been identified, their derivatives appearing in Eqs. \eqref{NS-linear}, \eqref{maurizio} and \eqref{maurizio2} can be easily computed, where the derivatives of $\Gamma^\alpha$ and $K^\alpha$ can be calculated explicitly from their expressions or numerically with finite difference method.



\subsection{Validation}
A common tool to solve a differential eigenvalue problem originating from hydrodynamic stability analysis is the Chebyshev-$\tau$ method \cite{canuto2007spectral}. Details about the implementation of the method and the advantage of using Chebyshev polynomials are overlooked  and the reader is referred to \cite{bourne2003hydrodynamic,arnone2023chebyshev}. Let us just summarise the main idea for implementation in case of two superposed fluids occupying a horizontal region. This problem is addressed in \cite{DongarraStraughWalker1996}, whose results are used to validate our in-house code.

For two superposed fluids, given that Chebyshev polynomials are naturally defined in $[-1,1]$, the numerical method is applied after a proper change of coordinates. Let us consider a two-dimensional domain where a  fluid, marked with $2$, is occupying the region $-1\leq y\leq 0$ underlying a different fluid denoted by $1$, bounded in $0\leq y\leq m$. In each region, the Orr-Sommerfeld equations hold and boundary conditions and interfacial conditions are appended to them. The differential eigenvalue problem to be solved is given by Eqs. 6.4-6.7 in \cite{DongarraStraughWalker1996}.

In order to employ the Chebyshev-$\tau$ method in such a problem, the following change of coordinates is performed
\begin{equation}
    z_1 = -2y-1 \quad z_2 =\frac{2}{m}y-1
\end{equation}
\begin{figure}[t]
    \centering    \includegraphics[scale=0.6]{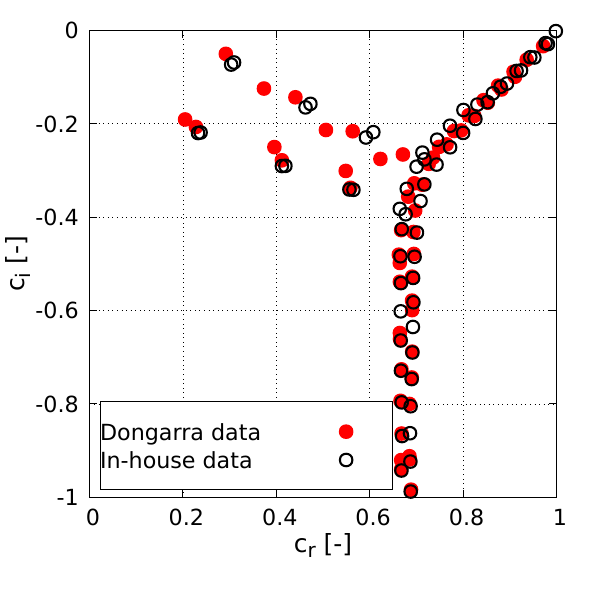}
    \caption{Comparison between the spectrum obtained in Figure 7 in \cite{DongarraStraughWalker1996} and the numerical results obtained with the in-house code.}    \label{fig:ChebyComparison}
\end{figure}
This change the coordinates maps, respectively, $y\in (-1,0)$ and $y\in (0,m)$ into $(-1,1)$, with the interface placed at $z_1=z_2=-1$ for both fluids. The derivatives have to be rewritten in the following way
\begin{equation}
\begin{aligned}
    & \frac{\text{d}}{\text{d}y}=-2\frac{\text{d}}{\text{d}z_1} \qquad y\in[-1,0]\\
    & \frac{\text{d}}{\text{d}y}=\frac{2}{m}\frac{\text{d}}{\text{d}z_2} \qquad y\in[0,m]
\end{aligned} 
\end{equation}
Hence, the differential eigenvalue problem can be written as
\begin{equation}
A\boldsymbol{\phi}=cB\boldsymbol{\phi}
\end{equation}
being $\boldsymbol{\phi}$ the eigenfunction, 
with
\begin{equation}
    A =\begin{pmatrix}
        A_1 & 0\\
        0 &  A_2\\
        BC_{inter,c}& BC_{inter,c}\\
        BC_{inter,0}& BC_{inter,0}\\
        BC_1 & 0\\
        0 & BC_2\\
    \end{pmatrix} \quad B = \begin{pmatrix}
        B_1 & 0\\
        0 &  B_2\\
        BC_{inter,c}& BC_{inter,c}\\
        0 & 0\\
        0 & 0\\
        0 & 0\\
    \end{pmatrix}
\end{equation}
where $A_i$ and $B_i$ are the operator of the left and right side, respectively, of the Orr-Sommerfeld equation for the fluid $i$. $BC_{inter,c}$ are the rows for the boundary conditions at the interface containing the eigenvalue $c$, $BC_{inter,0}$ are the rows for the boundary conditions at the interface not containing the eigenvalue and $BC_i$ are the boundary conditions involving only one of the two fluids. The boundary condition are used to make the matrix $B$ no longer singular in order to avoid the occurrence of spurious eigenvalues \cite{MCFADDEN1990228,arnone2023onset}. 

The in-house code has been tested and validated using the data presented in \cite{DongarraStraughWalker1996} as a benchmark and such a comparison is available in Figure \ref{fig:ChebyComparison} that shows coherence between the spectrum computed by the in-house code and the benchmark.

\subsection{Results and discussion}

\begin{figure}[t]
    \centering
    \includegraphics[scale=0.7]{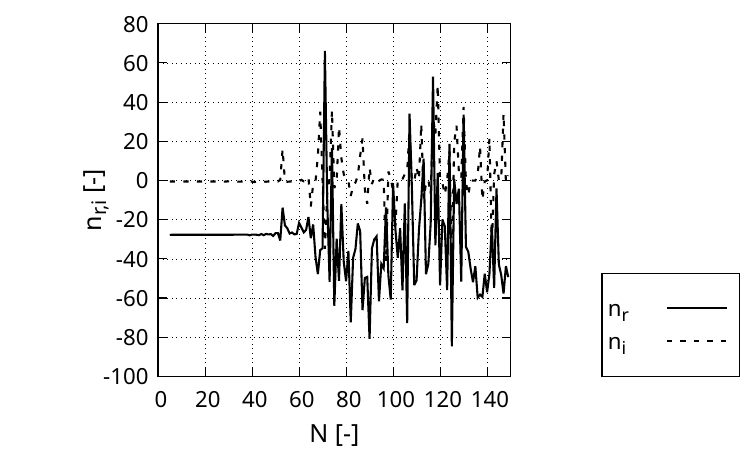}
    \caption{Value of the real and imaginary part of the leading eigenvalue as a function of $N$, with $T_{sat} = -15 ^\circ C, q = 30 kW/m^2, G =1200 kg/m^2, x = 0.7$ and $k=0.1$.}
    \label{fig:testN}
\end{figure}

The numerical method discussed in the previous subsection is used to solve the system made by the Eq. \eqref{NS-linear} and the boundary conditions \eqref{ICfirst}-\eqref{maurizio}-\eqref{maurizio2}, once a base flow profile has been chosen according to section \ref{subsection_baseflow}. 

Unlike the problem studied by \cite{DongarraStraughWalker1996}, the origin of the cartesian frame is set on the pipe wall (see Figure \ref{fig2}). Therefore, in order to recover the setting of \cite{DongarraStraughWalker1996},  following transformation has to be applied
\begin{equation}\label{eq:ChangeOfVariable1}
    \tilde{y} = \frac{y}{\delta}-1
\end{equation}
In the new reference frame, the origin is placed on the liquid-vapour interface and the pipe wall is identified by $\tilde{y}=-1$.

Now, the transformation proposed in \cite{DongarraStraughWalker1996} can be performed for each fluid-occupied region. Hence, for the liquid-phase region
\begin{equation}\label{eq:ChangeOfVariable2}
\begin{split}
    &     \tilde{y}=-\frac{y^l+1}{2}, \qquad y^l\in[-1,1]\\
    & \frac{\text{d}}{\text{d}\tilde{y}}=-\frac{2}{\delta}\frac{\text{d}}{\text{d}y^l}
\end{split}
\end{equation}
while for the vapour-phase region
\begin{equation}\label{eq:ChangeOfVariable3}
\begin{split}
    & \tilde{y}=\frac{m\left(y^v+1\right)}{2}, \qquad y^v\in[-1,1]\\
    & \frac{\text{d}}{\text{d}\tilde{y}}=\frac{2}{m\delta}\frac{\text{d}}{\text{d} y^v}
\end{split}
\end{equation}
where  $m=\frac{1}{2\delta}-1$.
Now, let us enumerate the steps followed to solve the stability problem, once $T_{sat}, G, q$ and $k$ are fixed:
\begin{enumerate}
    \item The base flow and all its physical quantities (coefficients of both phases velocities, void fraction, Boiling number, etc.) are computed
    in each point along the pipe.
    \item For a prescribed value of the vapour quality $x$, and of the corresponding $\delta (z)$, the aforementioned change of reference is performed.
    \item The differential eigenvalue problem is written in a matrix form $A\mathbf{x}=nB\mathbf{x}$ by virtue of the Chebyshev-$\tau$ method using $N$ polynomials.
    \item The algebraic eigenvalue problem is solved and the sign of the real part of the most unstable eigenvalue $n_r$ is studied: if $n_r > 0$ the liquid-vapour interface is disrupted, triggering dryout.
    \item The procedure continues from point 2 for a different value of vapour quality.
\end{enumerate}

The generalized eigenvalue problem $A\mathbf{x}=nB\mathbf{x}$ has the following form, using the notation shown in Section \ref{sec5}:
\begin{equation}\label{eq:GeneralEQ}
     \left(\begin{array}{cc}
		\mathcal{X}	& 0\\
        0&\mathcal{Y} \\
        \mathcal{A}^{1} & \mathcal{B}^{1} \\
        \mathcal{A}^2 & \mathcal{B}^2 \\
        \mathcal{A}^3 & \mathcal{B}^3 \\
        D &-D\\
        0 & 1\\
        0 &D\\
        1 &0\\
        D &0
	\end{array}\right)\left(\begin{array}{c}
		\widetilde{w}^l\\
		\widetilde{w}^v
	\end{array}\right)=n\left(\begin{array}{cc}
		\mathcal{W}	& 0\\
        0&\mathcal{W}\\
        \mathcal{C}^{1} & \mathcal{D}^{1} \\
        \mathcal{C}^2 & \mathcal{D}^2 \\
        \mathcal{C}^3 & \mathcal{D}^3 \\
        0 &0\\
        0 & 0\\
        0 &0\\
        0 &0\\
        0 &0
	\end{array}\right)\left(\begin{array}{c}
		\widetilde{w}^l\\
		\widetilde{w}^v
	\end{array}\right)
\end{equation}
where matrices $A$ and $B$ are to be considered as algebraic matrices and not differential operators. Being $N$ the number of Chebyshev polynomials employed for discretization, dimension of $A$ and $B$ in Eq. \eqref{eq:GeneralEQ} is $(2N+8)\times(2N+8)$.
Matrix $B$ is made non-singular by following a standard procedure via the last five rows, reducing matrices dimension to $(2N+3)\times(2N+3)$.

\begin{figure}[t]
    \centering
    \includegraphics[scale=0.7]{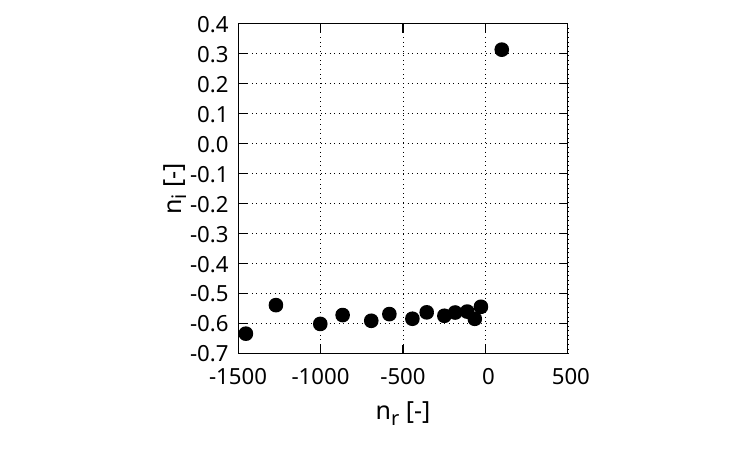}
    \caption{Spectrum obtained with Chebyshev-$\tau$ method with $N=20$ polynomials, for $T_{sat} = -15^\circ C, q = 30 kW/m^2, G =1200 kg/m^2, x = 0.7$ and $k=0.1$.}
    \label{fig:spectrum}
\end{figure}

The in-house code has been tested by varying the number of Chebyshev polynomials $N$. Figure \ref{fig:testN} shows one of the test cases. As usually happens for the Chebyshev-$\tau$ method, the eigenvalue converges when $N$ is approximately in the range $(12,40)$. Indeed, results from the numerical method oscillate when $N$ is large. This numerical phenomenon is due to the truncation error done when approximating the eigenfunctions with a fixed number of Chebyshev polynomials. In particular, the dimension of the matrix operators increases as $N^2$ and their entries become notably large exceeding the precision of the computational unit \cite{DongarraStraughWalker1996}. In the present study, $N=20$ has been chosen.
 
For a prescribed set of parameters, running the code to solve the eigenvalue problem will provide the spectrum as in Figure \ref{fig:spectrum}. For this case, the leading eigenvalue has a positive real part, showing that the interface is unstable.

\begin{figure}
    \centering
    \includegraphics[scale=0.3]{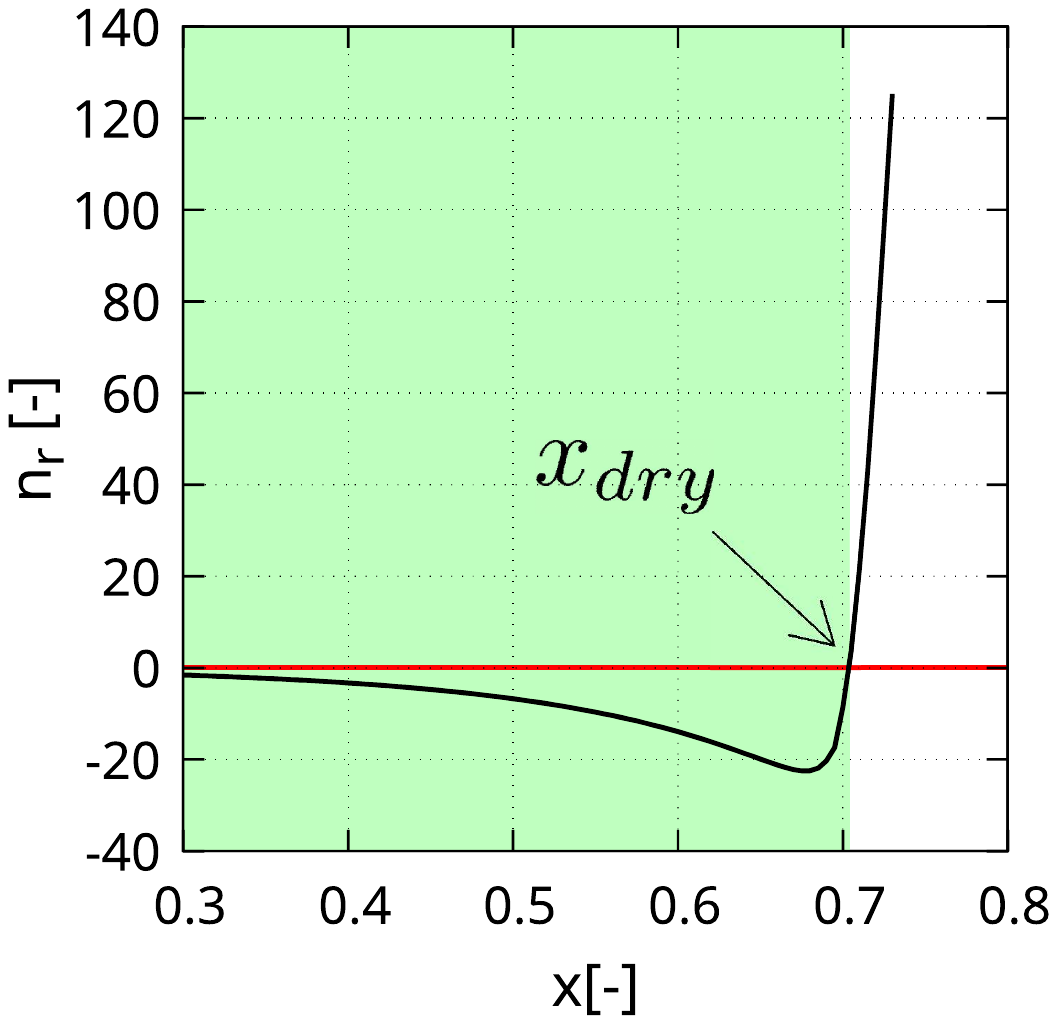}
    \caption{Plot of the real part of the most unstable eigenvalue for $T_{sat} = -15 ^\circ C, G = 2000 kg/(m^2s)$ and $q=30 kW/m^2$. The green shaded area represents the condition of stable vapour quality where the dryout does not take place.}
    \label{fig:eigenvap}
\end{figure}

Once the procedure has been iterated over a range of $x$ and the real part of the leading eigenvalue at each step is computed and stored, results of the procedure can be displayed like in Figure \ref{fig:eigenvap}. The most unstable eigenvalue has a negative real part until a certain value of vapour quality is reached. This value marks the transition between a stable and an unstable interface and it is named \emph{the dryout vapour quality} $x_{dry}$. Therefore, when the vapour quality $x$ exceeds $x_{dry}$, the liquid-vapour interface is unstable and dryout inception takes place.

\begin{figure}
    \centering
    \includegraphics[scale=0.8]{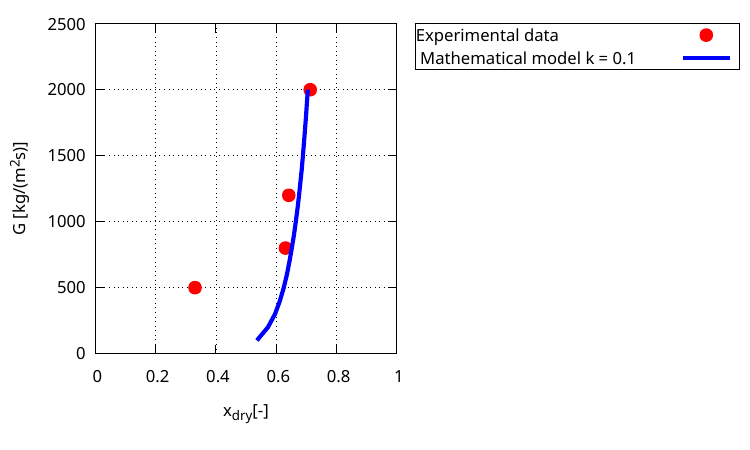}
    \caption{{Comparison between experimental data and numerical results for $T_{sat} = -15 ^\circ C$ and $q=30 kW/m^2$.}}
    \label{eigenflowmap}
\end{figure}

In Figure \ref{eigenflowmap}, the behaviour of the dryout vapour quality $x_{dry}$ is reported as a function of the mass flux $G$. The authors in \cite{cantini2025inception} have shown that in their experiments, increasing the inlet mass flux within the millichannel has a delaying effect of the occurrence of dryout. From our perspective this should result in a stabilising effect of $G$ on the liquid-vapour interface. The numerical procedure explained earlier in the paper has been repeated for a set of prescribed values of the mass flux $G$. Our numerical results show good agreement with experimental data, as shown in Figure \ref{eigenflowmap}. Hence, the trend of $\delta^+$ regime has been captured by the present model, with the only exception of the lowest value of the mass flux where experimental errors are larger and non-linear phenomena are triggered, making the linear analysis presented so far no longer suitable.
Let us also remark that the wavenumber $k$ has been chosen in such a way that the resulting $x_{dry}$ was as close as possible to the experimental value for $G=2000 kg/(m^2s)$.

\begin{figure}
    \centering
    \includegraphics[scale=0.8]{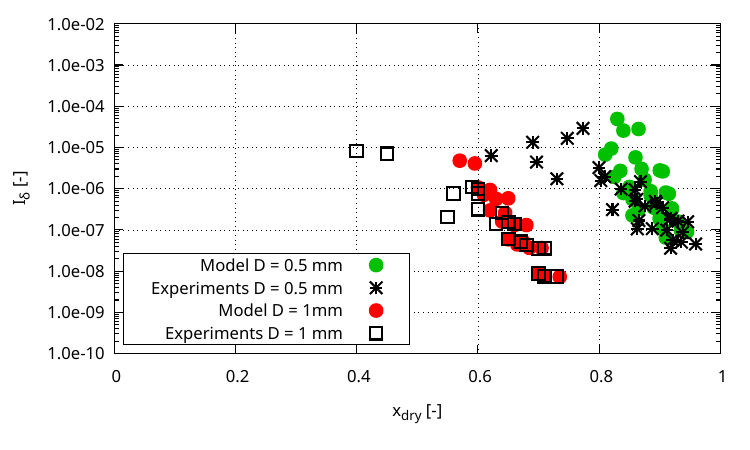}
    \caption{Comparison between results obtained via numerical investigations on the theoretical model and experimental data from two independent campaigns \cite{cantini2025inception,ducoulombier2011carbon}.
    Experimental data defining $I_{\delta^+}$ are plotted against the dryout vapour quality $x_{dry}$, determined both experimentally and numerically. 
    }
   \label{fig:ExpvsModel}
\end{figure}

Further, numerical results have been compared with the whole experimental dataset obtained during the campaign at CERN facilities for \cite{cantini2025inception}. The dryout instability factor embodies the compound effect of all the relevant parameters on the onset of dryout and it can be computed from the set of parameters available to run numerical investigations. As a result, once the dryout vapour quality is obtained from the linear stability analysis, it can be plotted as a function of $I_{\delta^+}$. Figure \ref{fig:ExpvsModel} summarises the results, showing again very good agreement of numerical results with experimental data in the tested range of parameters. Our model has been tested also for a different configuration in the $\delta^+$ regime. This configuration is the object of the research paper \cite{ducoulombier2011carbon} and involves a millichannel of diameter $D=0.5$mm. The agreement between numerical results and experimental data strengthen the validity of the approach taken in the present research paper.

In conclusion, the numerical predictions, along with the theoretical framework developed here, support the proposed physical mechanism whereby the dryout inception in millichannels for CO\textsubscript{2} in the $\delta^+$ regime is a consequence of the disruption of the interface between liquid and vapour phases, and is therefore triggered by interfacial instability.



\bibliographystyle{ieeetr}

\bibliography{mybib}

@book{ruggeri2015rational,
  title={Rational extended thermodynamics beyond the monatomic gas},
  author={Ruggeri, T. and Sugiyama, M.},
  year={2015},
  publisher={Springer}
}

@book{canuto2007spectral,
  title={Spectral methods: evolution to complex geometries and applications to fluid dynamics},
  author={Canuto, C. and Quarteroni, A. and Hussaini, M. Y. and Zang Jr, T. A.},
  publisher={Springer}
}

@article{bourne2003hydrodynamic,
  title={{Hydrodynamic stability, the Chebyshev tau method and spurious eigenvalues}},
  author={Bourne, D.},
  journal={Continuum Mechanics and Thermodynamics},
  volume={15},
  number={6},
  pages={571--579},
  year={2003},
  publisher={Springer}
}

@article{MCFADDEN1990228,
title = {{Elimination of spurious eigenvalues in the Chebyshev tau spectral method}},
journal = {Journal of Computational Physics},
volume = {91},
number = {1},
pages = {228-239},
year = {1990},
issn = {0021-9991},
doi = {https://doi.org/10.1016/0021-9991(90)90012-P},
url = {https://www.sciencedirect.com/science/article/pii/002199919090012P},
author = {G.B McFadden and B.T Murray and R.F Boisvert}
}

@article{arnone2023chebyshev,
  title={Chebyshev-$\tau$ method for certain generalized eigenvalue problems occurring in hydrodynamics: a concise survey},
  author={Arnone, Giuseppe and Gianfrani, Jacopo A and Massa, Giuliana},
  journal={The European Physical Journal Plus},
  volume={138},
  number={3},
  pages={281},
  year={2023},
  publisher={Springer}
}

@article{Cern,
title = {Co2 cooling is getting hot in high-energy physics},
journal = {CERN Courier},
year = {2012},
url = {https:
//cerncourier.com/a/co2-cooling-is-getting-hot-in-high-energy-physics/},
author = {Verlaart, B.},
}

@inproceedings{weislogel1998hydrodynamic,
  title={Hydrodynamic dryout in two-phase flows: Observations of low bond number systems},
  author={Weislogel, Mark M and McQuillen, John B},
  booktitle={AIP Conference Proceedings},
  volume={420},
  number={1},
  pages={413--421},
  year={1998},
  organization={American Institute of Physics}
}

@article{hsieh1972effects,
  title={Effects of heat and mass transfer on {R}ayleigh-{T}aylor instability},
  author={Hsieh, D. Y.},
  year={1972}
}

@article{cantini2025inception,
title = {{Inception of evaporative dryout for CO2 in milliscale pipe flows}},
journal = {International Journal of Heat and Mass Transfer},
volume = {251},
pages = {127299},
year = {2025},
issn = {0017-9310},
author = {Giulio Cantini and Desiree Hellenschmidt and Camila Pedano and Paolo Petagna and Carl Sangan and Mauro Carnevale}
}

@book{chandrasekhar1981hydrodynamic,
  title={Hydrodynamic and hydromagnetic stability},
  author={Chandrasekhar, Subrahmanyan},
  year={1981},
  publisher={Dover}
}

@misc{nist_webbook,
  author       = {{National Institute of Standards and Technology (NIST)}},
  title        = {NIST Chemistry WebBook, SRD 69},
  year         = {2025},
  doi          = {10.18434/T4D303},
  url          = {https://webbook.nist.gov/chemistry/}
}

@article{hsieh1978interfacial,
  title={Interfacial stability with mass and heat transfer},
  author={Hsieh, D Y},
  journal={The Physics of Fluids},
  volume={21},
  number={5},
  pages={745--748},
  year={1978},
  publisher={AIP Publishing}
}

@article{adham1995rayleigh,
  title={{The Rayleigh--Taylor and Kelvin--Helmholtz stability of a viscous liquid--vapor interface with heat and mass transfer}},
  author={Adham-Khodaparast, K and Kawaji, M and Antar, BN},
  journal={Physics of Fluids},
  volume={7},
  number={2},
  pages={359--364},
  year={1995},
  publisher={American Institute of Physics}
}

@article{gullo2017state,
  title={{State-of-the-art technologies for transcritical R744 refrigeration systems--a theoretical assessment of energy advantages for European food retail industry}},
  author={Gullo, P. and Tsamos, K. and Hafner, A. and Ge, Y. and Tassou, S.A.},
  journal={Energy Procedia},
  volume={123},
  pages={46--53},
  year={2017},
  publisher={Elsevier}
}

@article{cheng2008a,
title = {{New prediction methods for CO2 evaporation inside tubes: Part I – A two-phase flow pattern map and a flow pattern based phenomenological model for two-phase flow frictional pressure drops}},
journal = {International Journal of Heat and Mass Transfer},
volume = {51},
number = {1},
pages = {111-124},
year = {2008},
issn = {0017-9310},
doi = {https://doi.org/10.1016/j.ijheatmasstransfer.2007.04.002},
url = {https://www.sciencedirect.com/science/article/pii/S0017931007002839},
author = {Cheng, L. and G. Ribatski  and J. {Moreno Quibén} and  J.R. Thome},
}

@article{wojtan2005a,
title = {{Investigation of flow boiling in horizontal tubes: Part I—A new diabatic two-phase flow pattern map}},
journal = {International Journal of Heat and Mass Transfer},
volume = {48},
number = {14},
pages = {2955-2969},
year = {2005},
issn = {0017-9310},
doi = {https://doi.org/10.1016/j.ijheatmasstransfer.2004.12.012},
url = {https://www.sciencedirect.com/science/article/pii/S0017931005000268},
author = {L. Wojtan and T. Ursenbacher and J.R. Thome}
}

@article{hellenschmidt2021effects,
  title={Effects of saturation temperature on the boiling properties of carbon dioxide in small diameter pipes at low vapour quality: Heat transfer coefficient},
  author={D. Hellenschmidt and P. Petagna},
  journal={International Journal of Heat and Mass Transfer},
  volume={172},
  pages={121094},
  year={2021},
  publisher={Elsevier}
}

@book{collier1994,
  title={Convective Boiling and Condensation},
  author={Collier and Thome},
  year={1994},
  publisher={Oxford University Press}
}

@article{revellin2008conditions,
  title={Conditions of liquid film dryout during saturated flow boiling in microchannels},
  author={Revellin, R. and Haberschill, P. and Bonjour, J. and Thome, J.R.},
  journal={Chemical engineering science},
  volume={63},
  number={24},
  pages={5795--5801},
  year={2008},
  publisher={Elsevier}
}

@article{ducoulombier2011carbon,
  title={{Carbon dioxide flow boiling in a single microchannel--Part II: Heat transfer}},
  author={Ducoulombier, M. and Colasson, S. and Bonjour, J. and Haberschill, P.},
  journal={Experimental Thermal and Fluid Science},
  volume={35},
  number={4},
  pages={597--611},
  year={2011},
  publisher={Elsevier}
}

@article{YUN20032527,
title = {{Critical quality prediction for saturated flow boiling of CO2 in horizontal small diameter tubes}},
journal = {International Journal of Heat and Mass Transfer},
volume = {46},
number = {14},
pages = {2527-2535},
year = {2003},
issn = {0017-9310},
doi = {https://doi.org/10.1016/S0017-9310(03)00036-X},
url = {https://www.sciencedirect.com/science/article/pii/S001793100300036X},
author = {R. Yun and Y. Kim}
}

@book{drazin2004hydrodynamic,
  title={Hydrodynamic stability},
  author={Drazin, P.G. and Reid, W.H.},
  year={2004},
  publisher={Cambridge university press}
}

@book{hewitt2013annular,
  title={Annular two-phase flow},
  author={Hewitt, G.},
  year={2013},
  publisher={Elsevier}
}

@article{DongarraStraughWalker1996,
  title={Chebyshev tau-{QZ} algorithm methods for calculating spectra of hydrodynamic stability problems},
  author={Dongarra, J.J. and Straughan, B. and Walker, D.W.},
  journal={Appl. Numer. Math.},
  volume={22},
  pages={399--434},
  year={1996},
}

@article{arnone2023onset,
  title={The onset of penetrative convection in an inclined porous layer},
  author={Arnone, Giuseppe and Cantini, Giulio and Capone, Florinda and Carnevale, Mauro},
  journal={International Journal of Heat and Mass Transfer},
  volume={216},
  pages={124532},
  year={2023},
  publisher={Elsevier}
}

@book{bergman2011introduction,
  title={Introduction to heat transfer},
  author={Bergman, Theodore L and Lavine, Adrienne S and Incropera, Frank P and DeWitt, David P},
  year={2011},
  publisher={John Wiley \& Sons}
}

\end{document}